\documentclass[useAMS,usenatbib]{mn2e}
\usepackage{graphicx}
\usepackage{natbib}
\usepackage{amsmath}
\title[Analysis of the 2011 MOST light curve of TW~Hya]
{A stable quasi-periodic 4.18~d oscillation and mysterious occultations 
in the 2011 {\it MOST\/} light curve of TW~Hya.\thanks{Based on data 
from the {\it MOST\/} satellite, a Canadian Space Agency 
mission, jointly operated by Dynacon Inc., the University of Toronto 
Institute of Aerospace Studies, and the University of British Columbia, 
with the assistance of the University of Vienna.}
}
\author[M. Siwak et al.]
{Michal Siwak$^{1}$\thanks{E-mail: siwak@astro.as.up.krakow.pl},
Slavek M.\ Rucinski$^2$,
Jaymie M.\ Matthews$^3$,
David B.\ Guenther$^4$,\newauthor
Rainer Kuschnig$^{3,8}$,
Anthony F.\ J.\ Moffat$^5$,
Jason F.\ Rowe$^6$,
Dimitar Sasselov$^7$,\newauthor
Werner W.\ Weiss$^8$\\
$^1$Mount Suhora Astronomical Observatory, Cracov Pedagogical University,
ul.\ Podchorazych 2, 30-084 Krakow, Poland\\
$^2$Department of Astronomy and Astrophysics,
University of Toronto, 50 St.\ George St., Toronto,
Ontario, M5S~3H4, Canada\\
$^3$Department of Physics \& Astronomy, University of
British Columbia, 6224 Agricultural Road, Vancouver, B.C., V6T~1Z1, Canada\\
$^4$Institute for Computational Astrophysics,
Department of Astronomy and Physics,
Saint Marys University, Halifax, N.S., B3H~3C3, \\
Canada\\
$^5$D\'{e}partment de Physique, Universit\'{e}
de Montr\'{e}al, C.P.6128, Succursale: Centre-Ville,
Montr\'{e}al, QC, H3C~3J7, Canada\\
$^6$NASA Ames Research Center, Moffett Field, CA 94035, USA\\ 
$^7$Harvard-Smithsonian Center for Astrophysics,
60 Garden Street, Cambridge, MA 02138, USA\\
$^8$Universit\"{a}t Wien, Institut f\"{u}r Astronomie, 
T\"{u}rkenschanzstrasse 17, A-1180 Wien, Austria\\
}

\date{Accepted ;  Received ; in original form }

\pubyear{2014}

\begin{document}
\label{firstpage}
\maketitle

\begin{abstract}
We present an analysis of the 2011 photometric observations of TW~Hya
by the {\it MOST\/} satellite; this is the fourth continuous 
series of this type. 
The large-scale light variations are dominated by a strong,
quasi-periodic 4.18~d oscillation with superimposed, apparently chaotic 
flaring activity; the former is most likely produced by stellar rotation 
with one large hot spot created by a stable accretion funnel
while the latter 
may be produced by small hot spots, created at moderate latitudes 
by unstable accretion tongues. A new, previously unnoticed  
feature is a series of semi-periodic, well defined brightness dips 
of unknown nature of which 19 were observed 
during 43 days of our nearly-continuous observations. 
Re-analysis of the 2009 {\it MOST\/} light curve revealed 
the presence of 3 similar dips.
On the basis of recent theoretical results, we tentatively conclude that 
the dips may represent occultations of the small hot spots created 
by unstable accretion tongues by hypothetical optically thick clumps. 
\end{abstract}


\section{Introduction}
\label{intro}

Although originally considered a mysterious, isolated 
young K7Ve\footnote{or rather M2.5Ve from a more recent 
detailed estimate of \citealt{vacca2011}} 
star \citep{Herbig1978}, TW~Hya was later shown to be 
a genuine Classical T~Tauri-type Star (CTTS) \citep{ruc83}, 
one of two stars which still show vigorous accretion 
in a young (about 8 Myr) association now called TWA \citep{kastner97,barrado}. 
It is the closest ($56.4\pm7$~pc, \citealt{wichmann98}) T~Tauri-type star to us.

It is observationally well established that the accretion 
onto magnetic CTTSs occurs through the magnetospheric 
accretion mechanism, originally developed for accreting neutron stars \citep{ghosh77,ghosh79a,ghosh79b} and thereafter 
applied for CTTSs \citep{konigl91, cameron93, hartmann94, shu94}. 
Recent theoretical investigations 
\citep{romanowa04, romanowa08, kulkarni08, romanowa09, kulkarni09} 
of the magnetized-plasma accretion from innermost accretion disks 
are very relevant to the observational results presented in this paper. 
They suggest the following picture: For magnetospheres a few 
times the stellar radius in size (but no less than two stellar radii), 
the accretion from the surrounding disk can occur in either
a {\it stable}, a {\it moderately stable} or an {\it unstable\/} regime; 
The regime of accretion at a given time is controlled by the mass accretion rate 
and the disc viscosity parameter $\alpha$.
For small values of the viscosity parameter and a low accretion rate, 
the stable accretion takes the form of steady
plasma flows from the inner disc toward the stellar magnetic poles in two
funnels encircling the magnetosphere \citep{romanowa04}.
The funnels produce two antipodal banana-shaped 
hot spots which are almost unmovable on the star. 
Depending on the inclination angle and the misalignment angle 
between the stellar rotation axis and the magnetic pole, 
either one or both hot spots can be visible to an observer during a
single stellar rotation. 
The steady nature of the two accretion funnels results in two stable 
hot spots so that the flux changes should lead to 
fairly regular light curves with modulation corresponding to one (or a half) 
stellar rotation period \citep{romanowa04, kurosawa13}.

Increased disc viscosity and mass accretion rate may
lead to an onset of Rayleigh-Taylor (RT) instabilities in the inner accretion disc. 
The instabilities produce a few equatorial tongues, in which 
the matter is transferred directly from the disc onto the star. 
The matter hits the star at slightly lower, moderate latitudes 
and produces small hot spots \citep{romanowa08, kulkarni08, kulkarni09}. 
The stochastic behaviour of the tongues and the related 
hot spots results in progressively more chaotic synthetic light curves
as more spots are formed: While at 
the beginning of the RT instabilities the funnel component still produces 
a peak at the stellar rotation frequency (though not 
as steady as in the purely stable case), 
as time progresses the frequency spectrum starts to show increasingly 
more additional sporadic peaks produced by rotation into the view of 
multiple hot spots. This stage is called a {\it moderately stable} or 
an {\it intermediate accretion regime}.

In the fully {\it unstable regime}, for high values of $\alpha$ 
and for the mass accretion rate higher by an order of magnitude 
than in the {\it stable regime}, the hot spots are created by only 
a few tongues rotating around 
the star with angular velocity of the inner disc. It should be noted  
that in this regime the inner disc comes considerably 
closer to the star, as compared to its value during the {\it stable regime} 
(see Sec. 3.1 in \citealt{kulkarni08}). 
The shape, intensity, number and position of the hot spots 
change in the inner disc dynamical timescale, the stellar rotation 
frequency is no longer visible in the frequency spectrum 
and the synthetic light curves, Fourier and wavelet spectra are very chaotic. 
Because the tongues move at approximately the inner-disc 
orbital frequency, the hot spots no longer co-rotate with the star  
but move in relation to the photosphere. 
From time to time one or more of the tongues produce a hot spot or a group
of spots, which may dominate the overall light changes for a short time -- 
a process leading to drifting 
quasi-periodic light variations reflecting the inner disc keplerian 
frequency \citep{kulkarni08, romanowa09, kulkarni09}.

Our {\it MOST\/} photometric observations of TW~Hya 
started in the 2007 and 2008 seasons 
\citep{ruc08} and continued through the 2009 season \citep{siwak11}. 
During the three seasons, the star showed apparently an
irregular behaviour, with flicker-noise characteristics in the Fourier spectrum. 
Although the star is visible nearly pole-on 
($i\approx15$~deg; see \citealt{ruc08} for previous 
estimates and \citealt{donati11}), 
no single periodicity dominated even when the observed variations reached
about 0.5~mag, occasionally even as much as 1~mag. 
The long series of observations in 2008 and 2009, 
analyzed with the wavelet technique, led to an unexpected 
and significant result: We have firmly established the presence of 
oscillatory variations which appeared in the accessible range of 
about 9 to 1.3 days and shortened their periods 
by typically a factor of two within a few weeks. 
We originally interpreted this phenomenon as caused 
by hypothetical hot plasma condensations spirally revolving 
in the inner accretion disc toward its inner radius. 
However, these variations could be better interpreted as
caused by hot spots produced during the {\it unstable regime} 
of accretion\footnote{The presence of such hot spots 
at moderate and even low stellar latitudes is confirmed 
by spectroscopic observations obtained in March 2008 
by \cite{donati11}, i.e.\ exactly during the 2008 {\it MOST\/} 
observations. The {\it MOST\/} revealed that the light curve 
and the wavelet spectrum was then dominated by a quasi-periodic feature drifting 
in its quasi-period from $\sim4$ to $\sim3$~days 
(see in Fig.8 in \citealt{ruc08}).}. 
This conclusion finds support in relatively 
low inner disc temperature (1100-1400~K, 
\citealt{eisner06}) and a blue colour index of these 
oscillatory variations which is in qualitative accordance 
with $T\sim 8000$~K of the variable hot source (see \citealt{batalha02}). 
Due to the absence of any stable periodicity in the 
2008 and 2009 {\it MOST\/} light curves,  
we infer that the strongly {\it unstable regime} of accretion 
-- solely through fast moving tongues created by RT instabilities -- 
operated in TW~Hya at the time. 

According to \cite{romanowa08}, episodes of stable and unstable accretion may alternate
depending on the accretion rate. We describe here observations of the first clear 
instance of the {\it moderately stable accretion regime} in TW~Hya  
which apparently took place in 2011 and was observed
by {\it MOST\/} during the fourth series of observations which lasted 
over 43 days (Section~\ref{observations}). 
We discuss the new results obtained using Fourier and 
wavelet analyses of these observations in Section~\ref{results}. 
The advantage of a better temporal coverage is that it permitted discovery of
 well defined drops in the star's brightness 
(Section~\ref{eclipses}). 
A summary of the results is given in Section~\ref{summary}.

\begin{figure*}
\includegraphics[height=180mm,angle=-90]{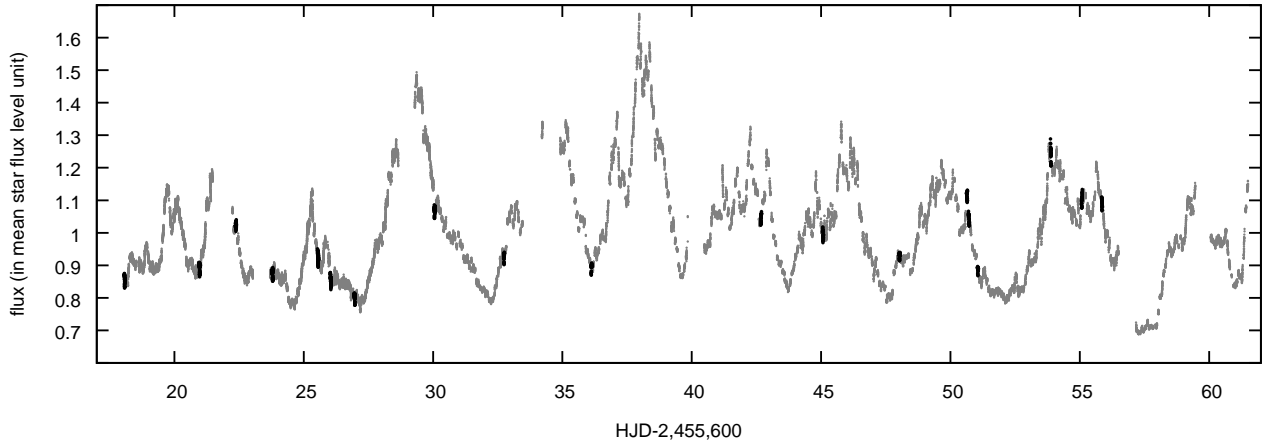}\\
\caption{The 2011 light curve of TW~Hya in flux units, scaled 
to unity at the mean brightness level. All individual data points are shown. 
Observations obtained during {\it MOST\/} orbits which showed the ``occultations'' 
(see Section~\ref{eclipses}) are represented by darker points.
}
\label{Fig.1}
\end{figure*}

\section{Observations and data reductions}
\label{observations}

The optical system of the {\it MOST\/} satellite consists of a 
Rumak-Maksutov f/6, 15~cm reflecting telescope. 
The custom broad-band filter covers the spectral range of 
350 -- 700~nm with the effective wavelength located close to the Johnson $V$ band.
The pre-launch characteristics of the mission are described by 
\citet{WM2003} and the initial post-launch performance by \citet{M2004}.

The fourth run of nearly continuous TW~Hya observations,
utilizing the {\it direct-imaging} data-acquisition mode, took place
over 43.33~days between 24 February and 11 April, 2011, during 545 satellite 
orbits. 
Because the star is not in the Continuous Visibility Zone of the satellite 
and some short time-critical observations of other targets were done in parallel, 
the effective total time coverage was 16.92~d i.e. 39 percent of the run total length.

The individual exposures were 30~s during the first part of the run (12.16~days) 
and 60~s during the second part (31.17~days). For photometric reductions,
the {\it dark} and {\it flat} calibration frames were 
obtained by averaging a dozen {\it empty-field} images specifically taken during 
the 60~sec run, or -- for the case of the 30~sec long exposures -- from frames with 
the target localized far beyond its optimal position due to occasional satellite 
pointing errors.
Aperture photometry of the stars was obtained using the {\it dark} and {\it flat} 
corrected images by means of the {\small \sc DAOPHOT~II} package \citep{stet}.
As in our previous investigations, a weak correlation between the star flux and 
the sky background level within each {\it MOST\/} orbit was noted and removed; 
it was most probably caused by a small photometric nonlinearity in 
the electronic system.

We obtained a well defined light curve for the
whole duration of the observations (Fig.~\ref{Fig.1}). 
The typical error of a single data point is about 0.011~mag. 
The median value of error ($\sigma$) of 545 averaged 
points (formed for each satellite orbit of 101~min) 
is 0.0073, with the full range between 0.00014--0.044 in units of the mean 
normalized flux for the star. 
Such values of errors are obviously significantly increased due to the variability 
intrinsic to the star, occurring in time scales shorter then the length of 
a single {\it MOST\/} orbit.\\
In contrast to the previous {\it MOST\/} runs which generally showed erratic behaviour 
requiring further analysis to reveal regularities in temporal variations, the variations 
observed in 2011 were surprisingly regular showing roughly equidistant spikes at typical 
separation of about 4 days. This morphology is new to TW~Hya as is 
the quasi-period which was never observed before to be so persistent.

\begin{figure}
\includegraphics[width=60mm,angle=-90]{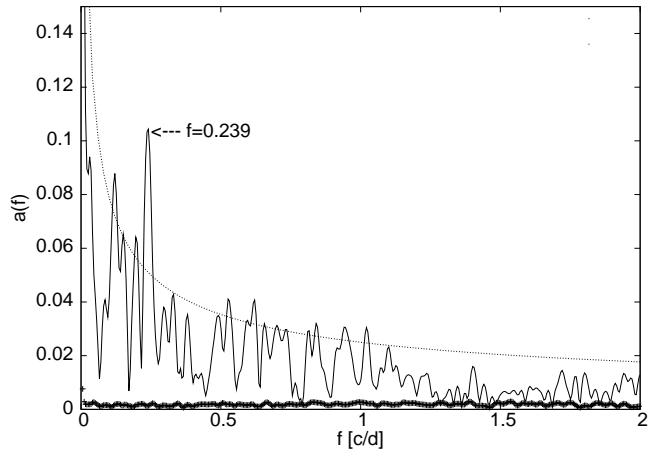}
\caption{The frequency $f$ spectrum of TW~Hya in cycles per day obtained from all 
2011 {\it MOST\/} observations (the continuos line). 
A thick black line along the horizontal axis represents errors of amplitude $a(f)$ 
obtained from the bootstrap sampling technique. 
The dotted line represents the shape of flickering noise: $a \propto 1/\sqrt{f}$  
shown here with arbitrary scaling.
}
\label{Fig.2}
\end{figure}

\begin{figure}
\includegraphics[width=85mm]{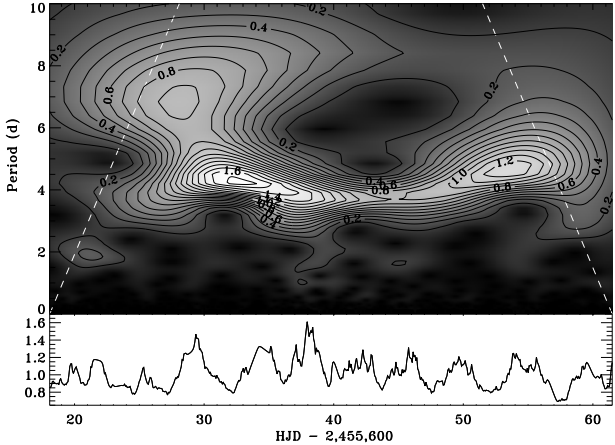}
\caption{The Morlet-6 wavelet transform of the 2011 TW~Hya {\it MOST\/} data. 
The amplitudes of the transform are expressed by grey scale intensities 
and contours. Edge effects are present outside the white 
broken lines but they do not affect our conclusions.
At the bottom, the {\it MOST\/} light curve (in mean star level flux units), 
re-sampled into a uniformly distributed time-grid with 0.07047~d spacing, is shown.}
\label{Fig.3}
\end{figure}

\section{Results of the light-curve analysis}
\label{results}

\subsection{Fourier analysis}
\label{fourier}

We performed analysis of the light curve in a similar way
to \citet{ruc08,ruc10} and \citet{siwak11}. 
The bootstrap sampling technique permitted evaluation of the mean standard
errors of the amplitude $a(f)$, where $f$ is the frequency.
We used for the Fourier analysis all 29,230 single data points. 

The amplitude spectrum has very similar characteristics to those described in our 
previous investigations, with dominant flicker-noise $a \propto 1/ \sqrt{f}$ 
characteristics (Fig.~\ref{Fig.2}).
Yet, there is one $f=0.239\pm0.014$~c/d significant peak, 
which -- on the basis of the regular 
variations visible in the full light curve as in Fig.~\ref{Fig.1}
-- one might be tempted to assign the $4.18\pm0.25$~day variations 
to rotation of the star. 
However, as we know from our previous investigations, TW~Hya can demonstrate 
an amazingly rich spectrum of temporal variations with quasi-periods 
ranging between 1.3 and 9 days.

\subsection{Wavelet analysis}
\label{wavelet}

To obtain uniform data sampling required for 
the wavelet analysis and to remove a few 
interruptions in the data acquisition (see Sec.~\ref{observations}), 
we interpolated the 545 mean satellite-orbit flux points 
into a grid of 617 equally-spaced points at 0.07047 day. 
As we found previously (\citealt{ruc08, ruc10, siwak11}),   
the Morlet-6 wavelet provided the best match between the time-integrated
power spectrum and original frequency spectrum of the star (Fig.~\ref{Fig.3}).

The new results are very different from those obtained during
the 2008 and 2009 {\it MOST\/} observations: \newline 
(1)~Starting from $HJD\approx2,455,627$ the light curve and the wavelet spectrum 
is dominated by one oscillation of 4.18~d which is strongly visible in the
Fourier spectrum in Fig.~\ref{Fig.2} at $f=0.239$~c/d.\newline 
(2)~The oscillation represented by this 4.18~d signal is fairly stable and it does 
not show any tendency for evolution towards shorter periods as was observed during 
the previous runs.\newline
(3)~Traces of the shortening tendencies may appear in other features in Fig.~\ref{Fig.3}, 
but they have low significance in comparison to the main, stable oscillation of about 
4.18 days.

We conclude that quasi-periodic oscillations with shortening periods do not always 
play a primary role in TW~Hya light variations, 
contrary to the previous wavelet analysis results. 
One should keep in mind that the statistics are convincing but not overwhelming: 
We had only 11 days of observations in 2007 which was too short
an interval to analyze with wavelets. 
The long observations of 2008 and 2009 extending over 46.7 days 
and 40.3 days showed the very clear trends to shorter periods. 
It is only the current, fourth run of over 43 days which does not confirm the tendency 
of the shortening quasi-periodicities which makes it particularly significant 
in view of recent theoretical investigations.

\subsection{Interpretation}

When interpreted in terms of the numerical simulations by \citet{romanowa08}, 
\citet{kulkarni08} and \citet{kurosawa13}, the events observed during the entire 
2011 {\it MOST\/} run occurred during a {\it moderately stable regime} of accretion 
onto TW~Hya. 
As was described in the Introduction, 
in this picture, the primary 4.18~d almost stable quasi-periodicity 
could be produced by rotation-modulation in visibility of
a banana-shaped, large hot spot created at the 
footprint of a steady accretion funnel striking the star 
close to its magnetic pole. The secondary peaks visible 
before $HJD\approx2,455,627$ (and other peaks overlapping with 
and modulating the primary 4.18~d signal) could be then caused 
by stochastic accretion tongues producing small hot spots 
at moderate stellar latitudes. The secondary spots, responsible 
for the drifting quasi-periodicities, were apparently playing 
a secondary role in the large scale light variations during the reported
here observations (see item no.~3 in Section~\ref{wavelet}). Thus,
within the picture presented by \citet{kulkarni08}, the clearly defined 
onset of accretion through the steady accretion funnel 
(at about $HJD\approx2,455,627$) led likely to a drop in the mass-accretion rate
by almost an order of magnitude in an interval as short as a week.

We note that \citet{batalha02} obtained a similar value of the stellar rotation period of
$4.4\pm0.4$~d from their veiling measurements obtained in May and July 1998 
and the archival Johnson $B$-filter photometry of the star. 
From changes of hot spot projected size, they inferred its extent in
latitude to be smaller than 20~deg. 
Assuming that the spot was a long-term feature created during the {\it stable} 
or {\it moderately stable regime} of accretion, 
their estimate is in conflict with the theoretical 
results of \citet{romanowa04} showing that it must be localized strictly close 
to the magnetic pole and the stellar rotation axis. 
The misalignment angle between rotation-axis and magnetic pole for TW~Hya seems 
to be smaller than 10~deg \citep{donati11} so that the apparent
movement of the spot and thus light variations should be small.  
A plausible solution of this discrepancy could be a scenario in which
the 4.4~d signal is due to a hot spot (or a group of spots) 
formed at moderate latitudes by long-term accretion tongues 
concentrated on one side of the star \citep{kulkarni08}. 
However, it is rather unlikely that the latest situation would be stable 
for over 2 months of \citet{batalha02} observations. 

The apparent attractiveness of the {\it moderately stable accretion regime} 
in explanation of the observed by us light variations in 2011 
and in 1998 by \citet{batalha02} may, however, be even more challenged. 
TW~Hya exhibits periodic 3.57~d radial velocity variations
which were detected spectroscopically by \cite{setiawan08} 
and then confirmed by \cite{huelamo08} and \cite{donati11}. 
The authors of the two recent publications attributed them 
to a high-latitude {\it cold spot\/} on the stellar photosphere, which 
remained permanently on the star for at least 2 years; its close vicinity 
to the major hot spot would then lead to a higher luminance contrast 
in the polar area of the star.
A combination of large hot and cold spots and the low rotation-axis
inclination of TW~Hya, with added possibilities
of strongly differential rotation open up possibilities to explain
the observed amplitudes of light variations and the discrepancy between the photometric 
period of 4.18 -- 4.4 days and the spectroscopic period of 3.57~d by adjustment 
of several parameters in the resulting complex geometry; we prefer to refrain 
from such an exercise for now.

\section{The occultations}
\label{eclipses}

\begin{figure*}
\includegraphics[height=160mm,angle=-90]{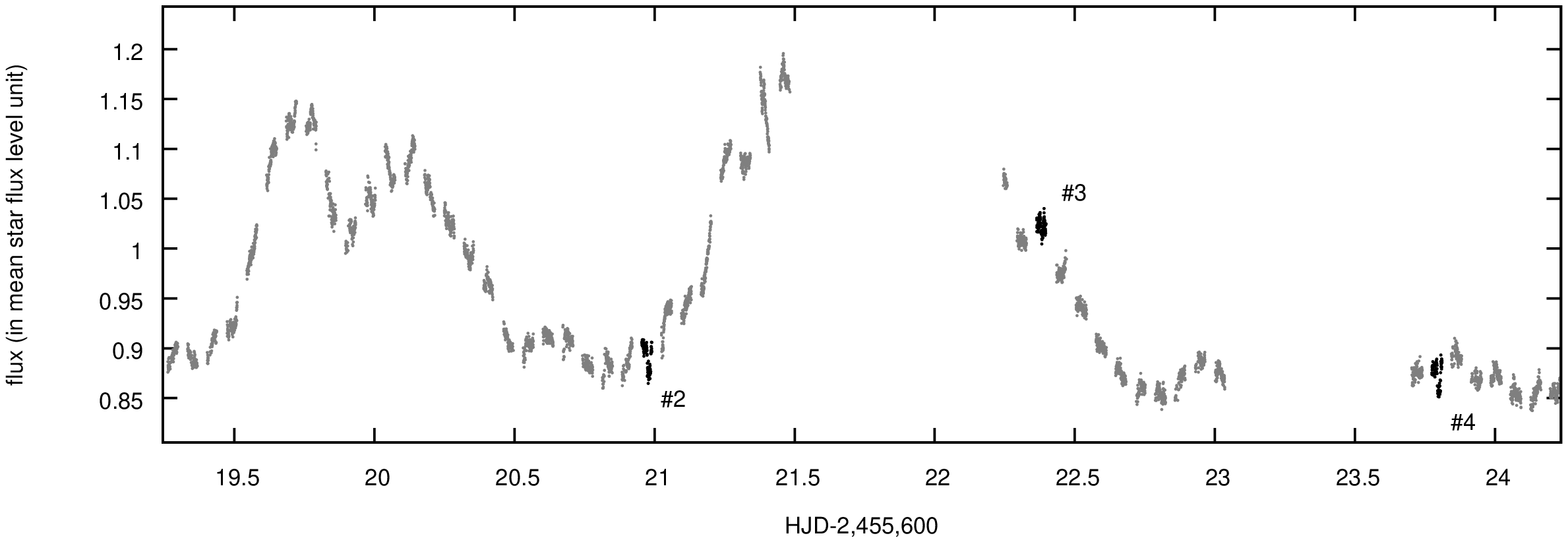}
\centerline{%
\begin{tabular}{c@{\hspace{0pt}}c@{\hspace{0pt}}c@{\hspace{0pt}}c}
\includegraphics[height=42mm,angle=-90]{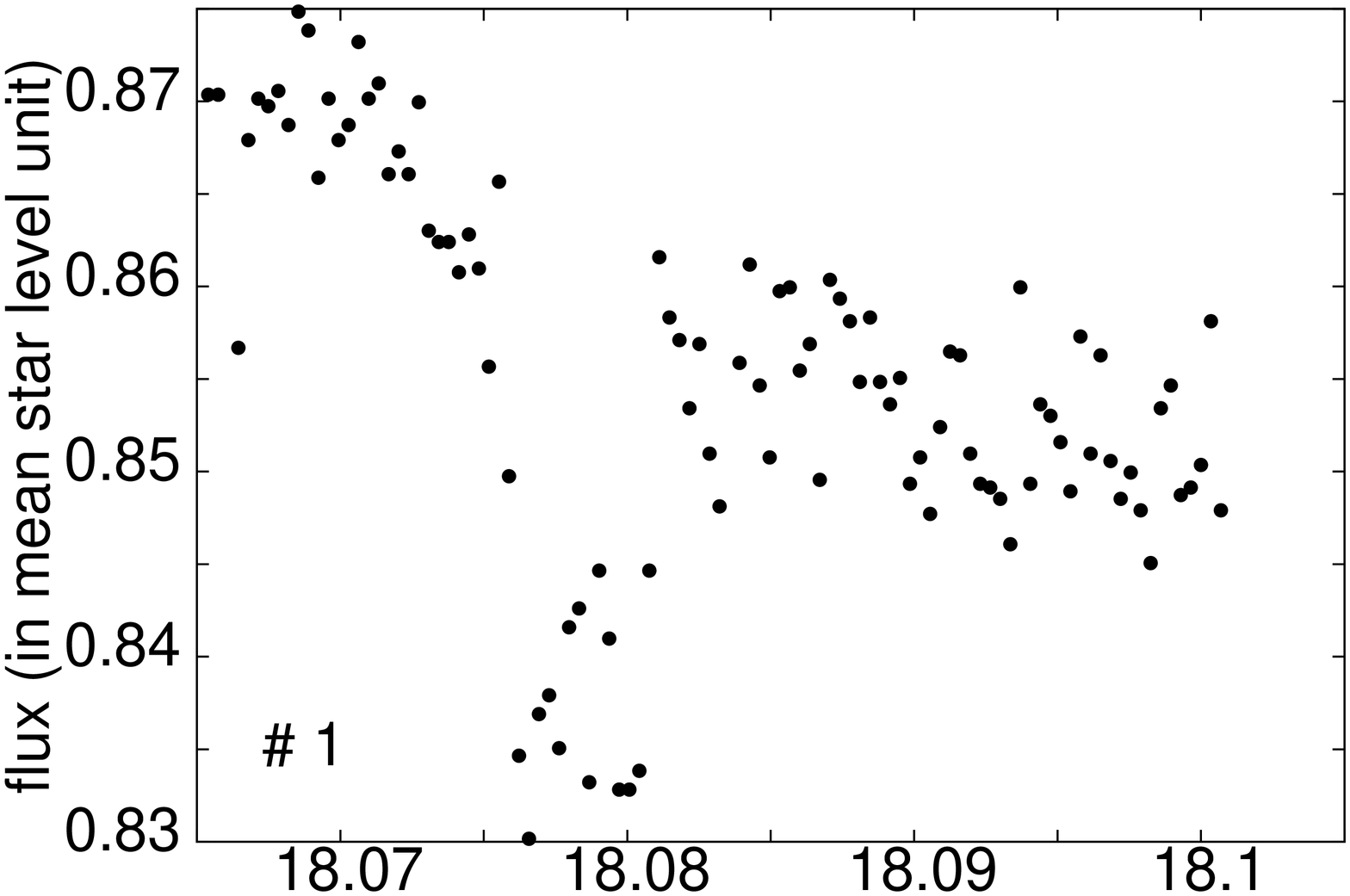}
\includegraphics[height=42mm,angle=-90]{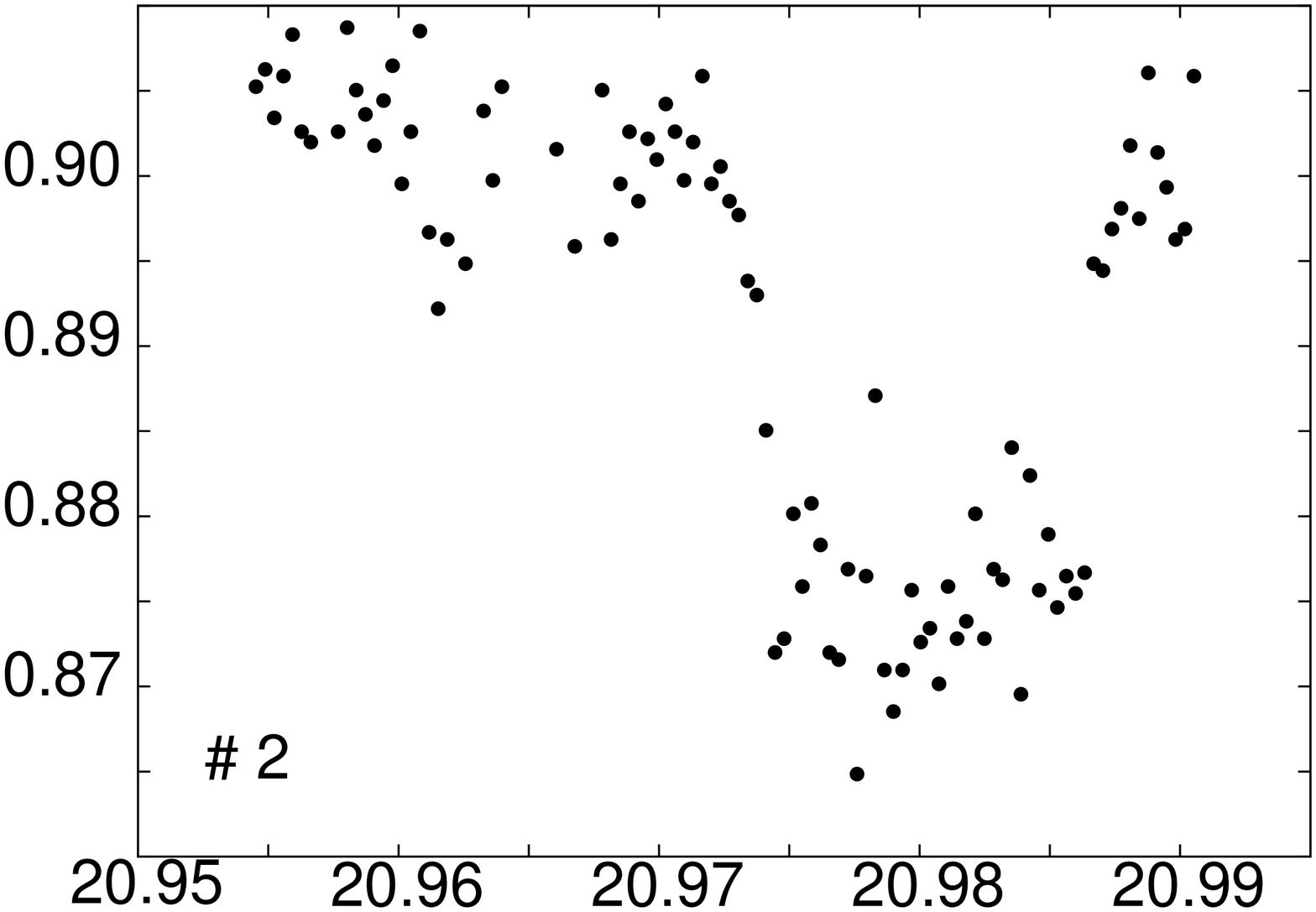}
\includegraphics[height=42mm,angle=-90]{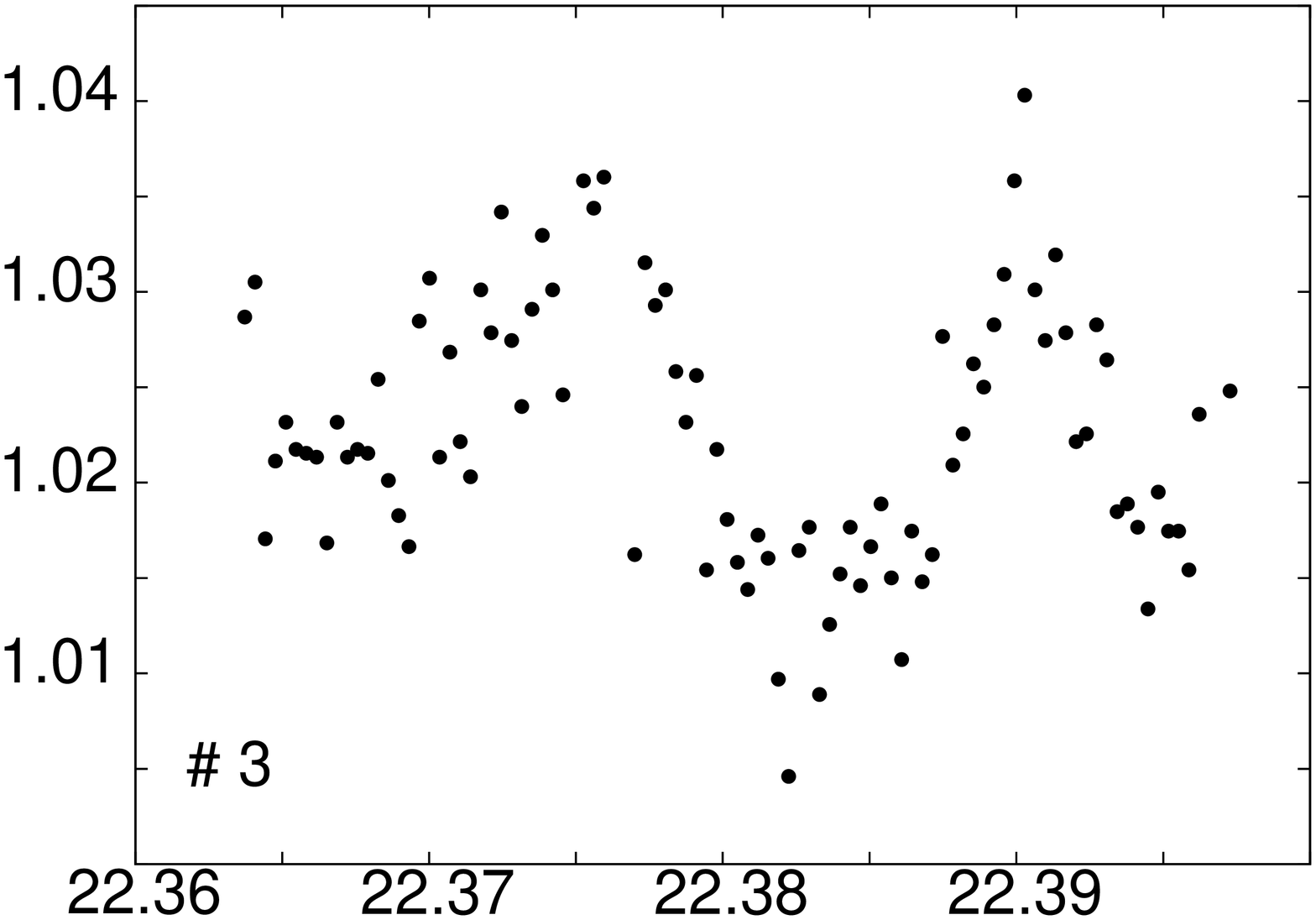}
\includegraphics[height=42mm,angle=-90]{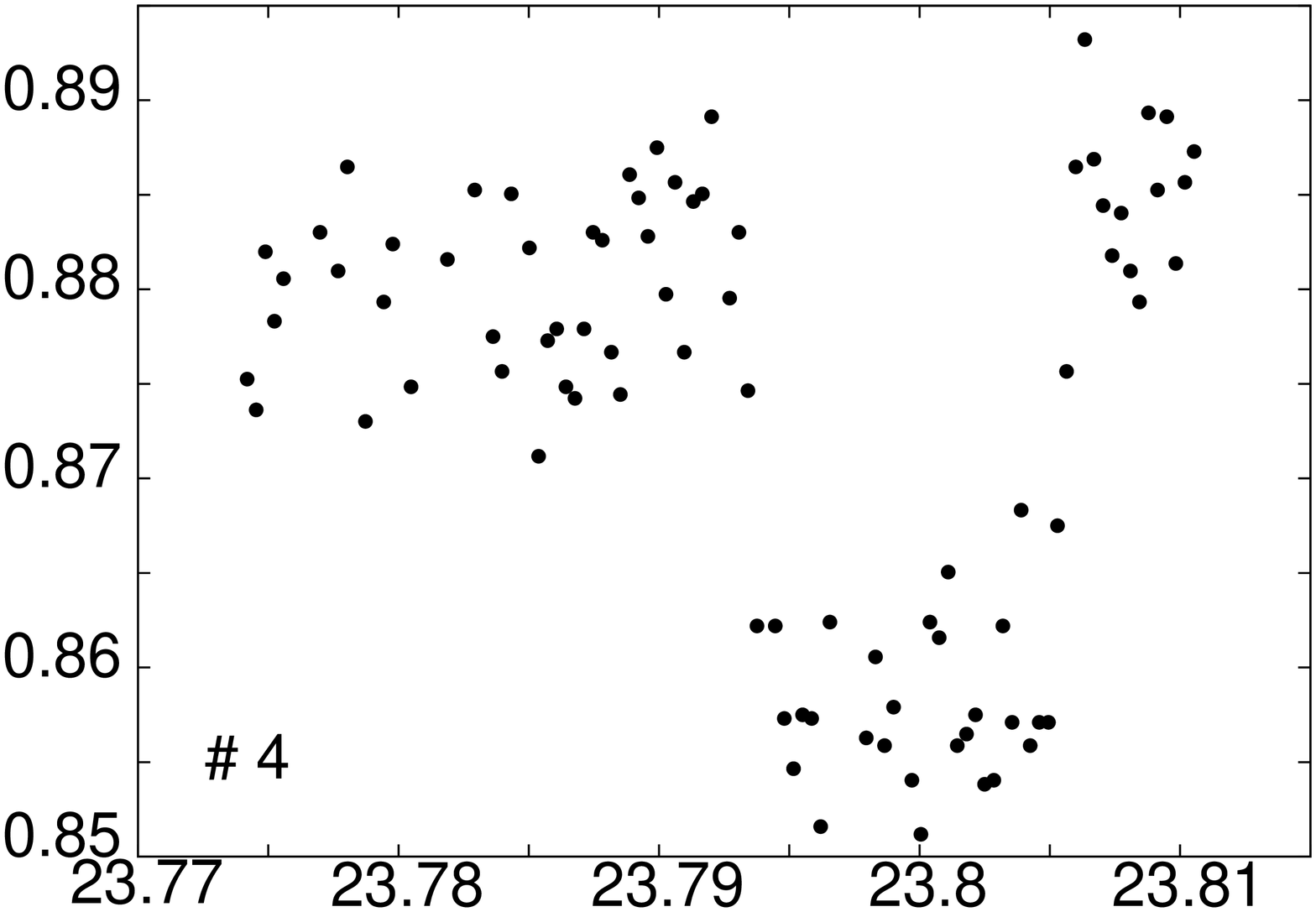}\\
\includegraphics[height=42mm,angle=-90]{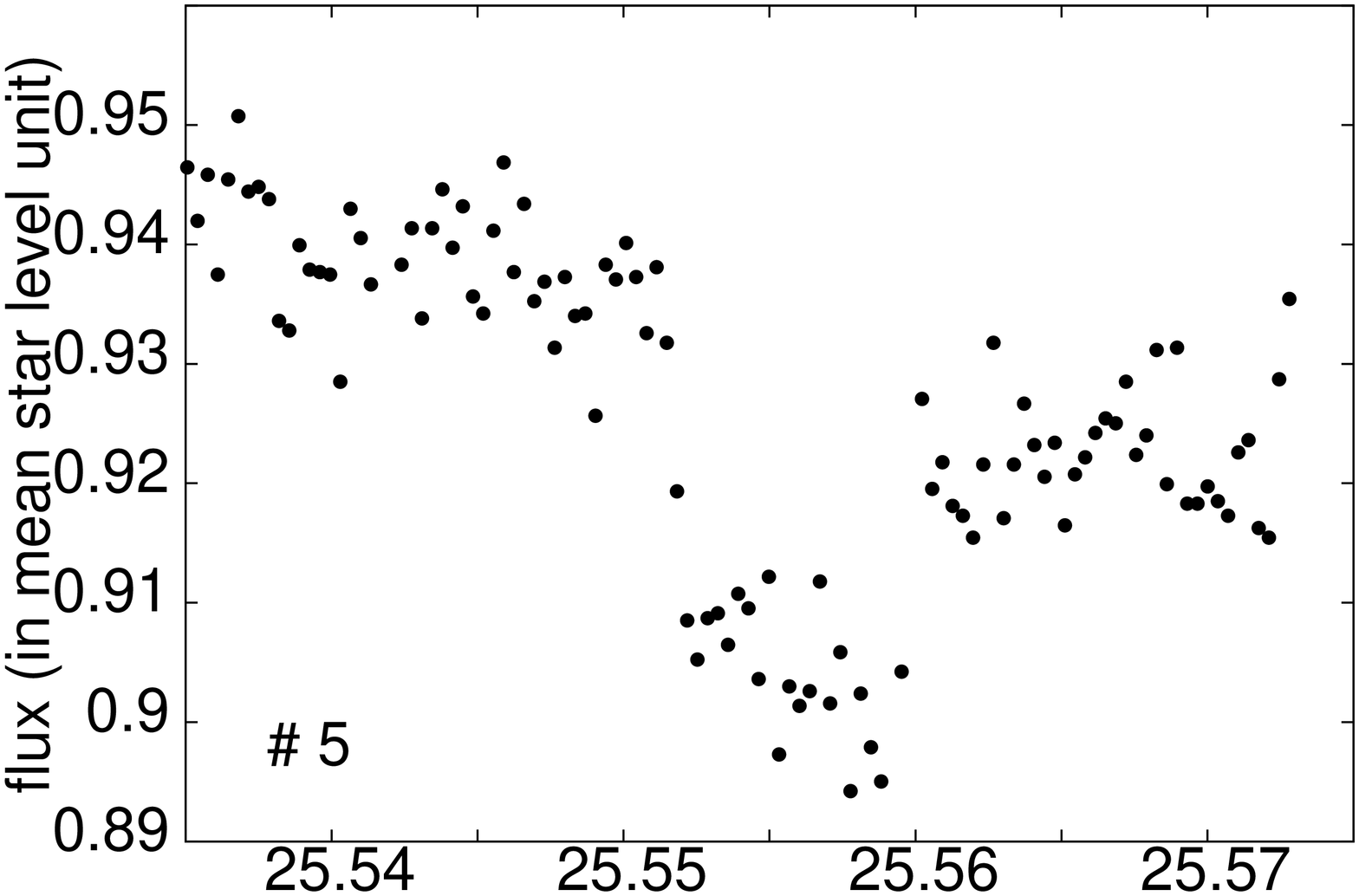}
\includegraphics[height=42mm,angle=-90]{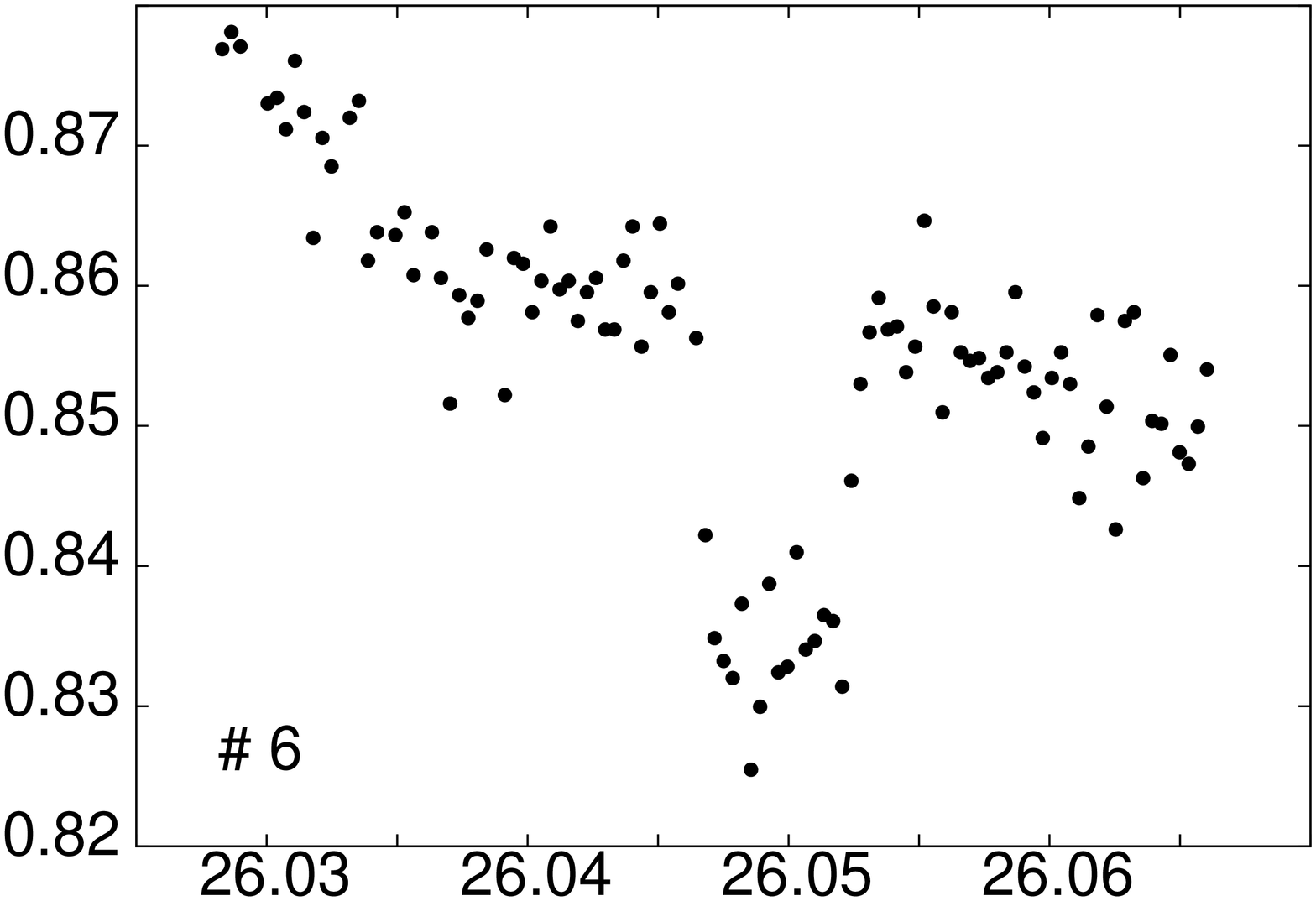}
\includegraphics[height=42mm,angle=-90]{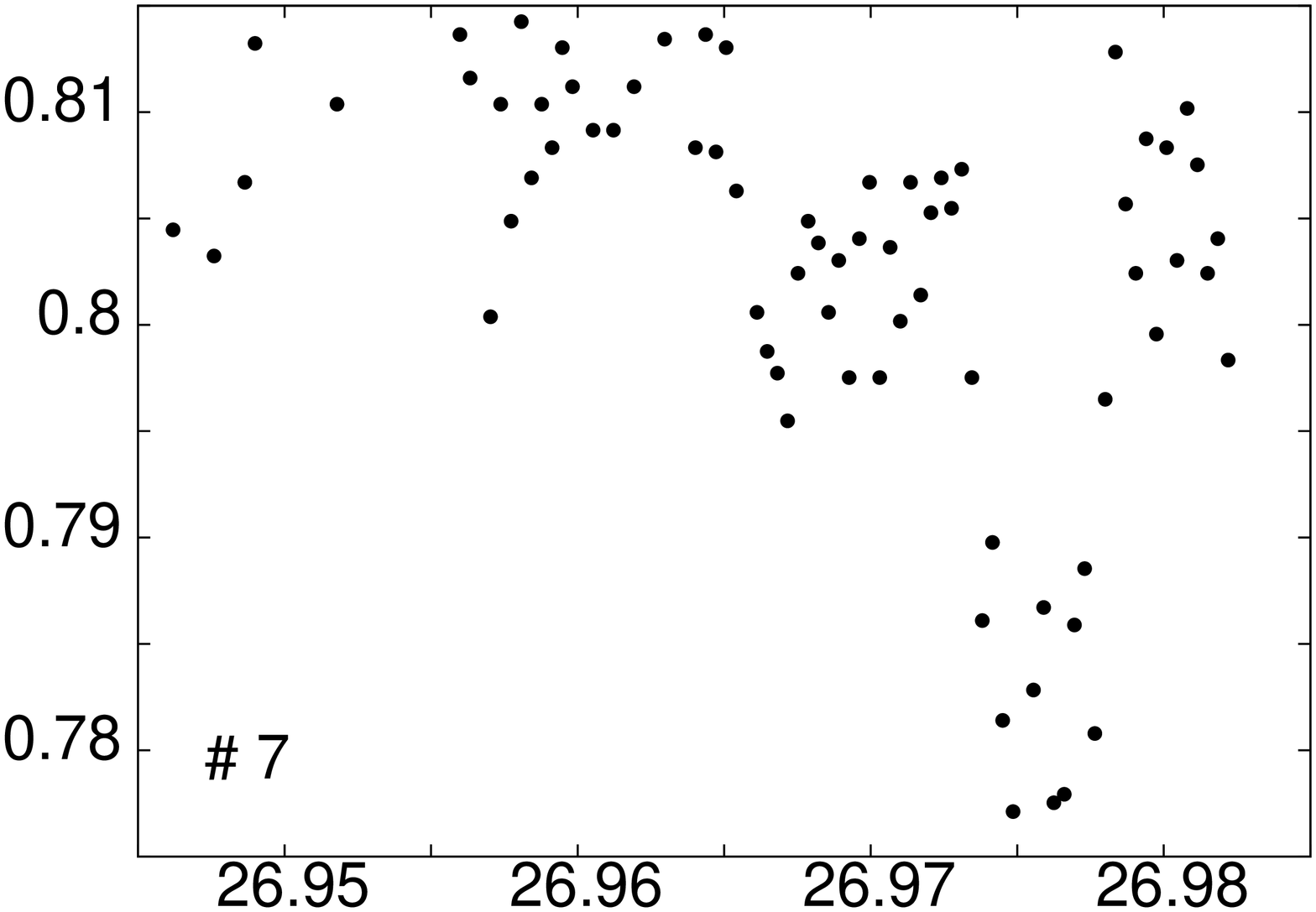}
\includegraphics[height=42mm,angle=-90]{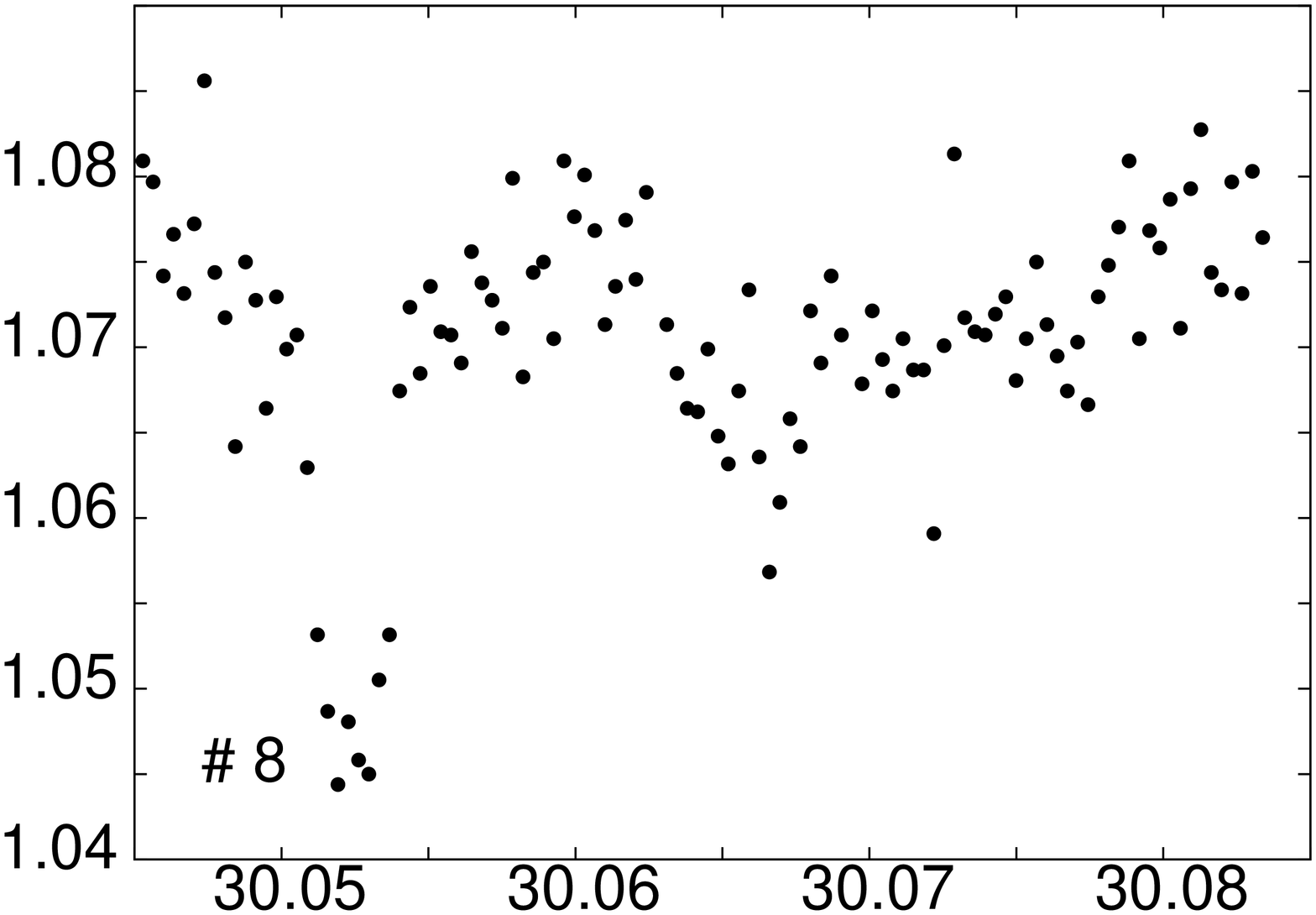}\\ 
\includegraphics[height=42mm,angle=-90]{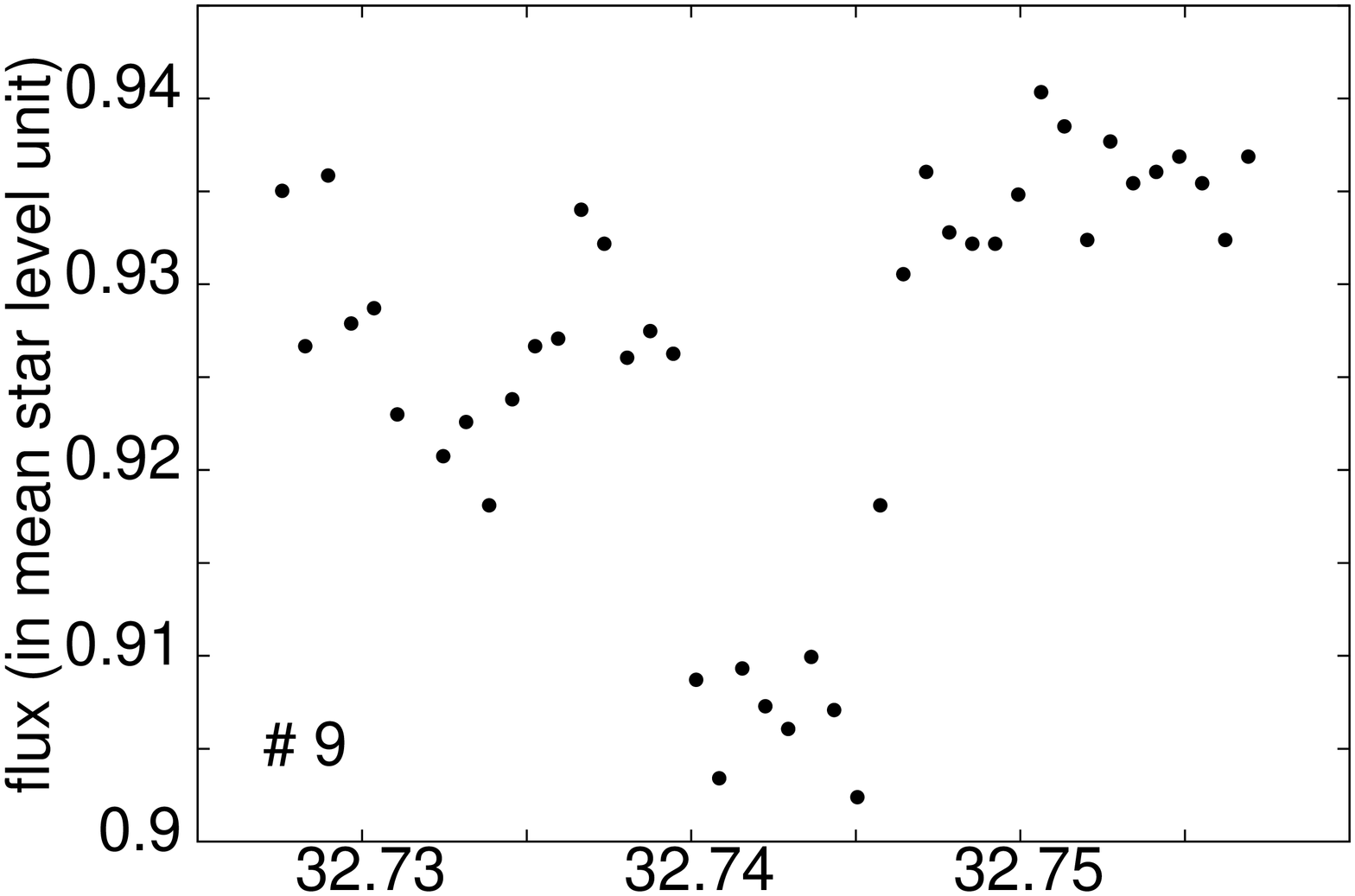} 
\includegraphics[height=42mm,angle=-90]{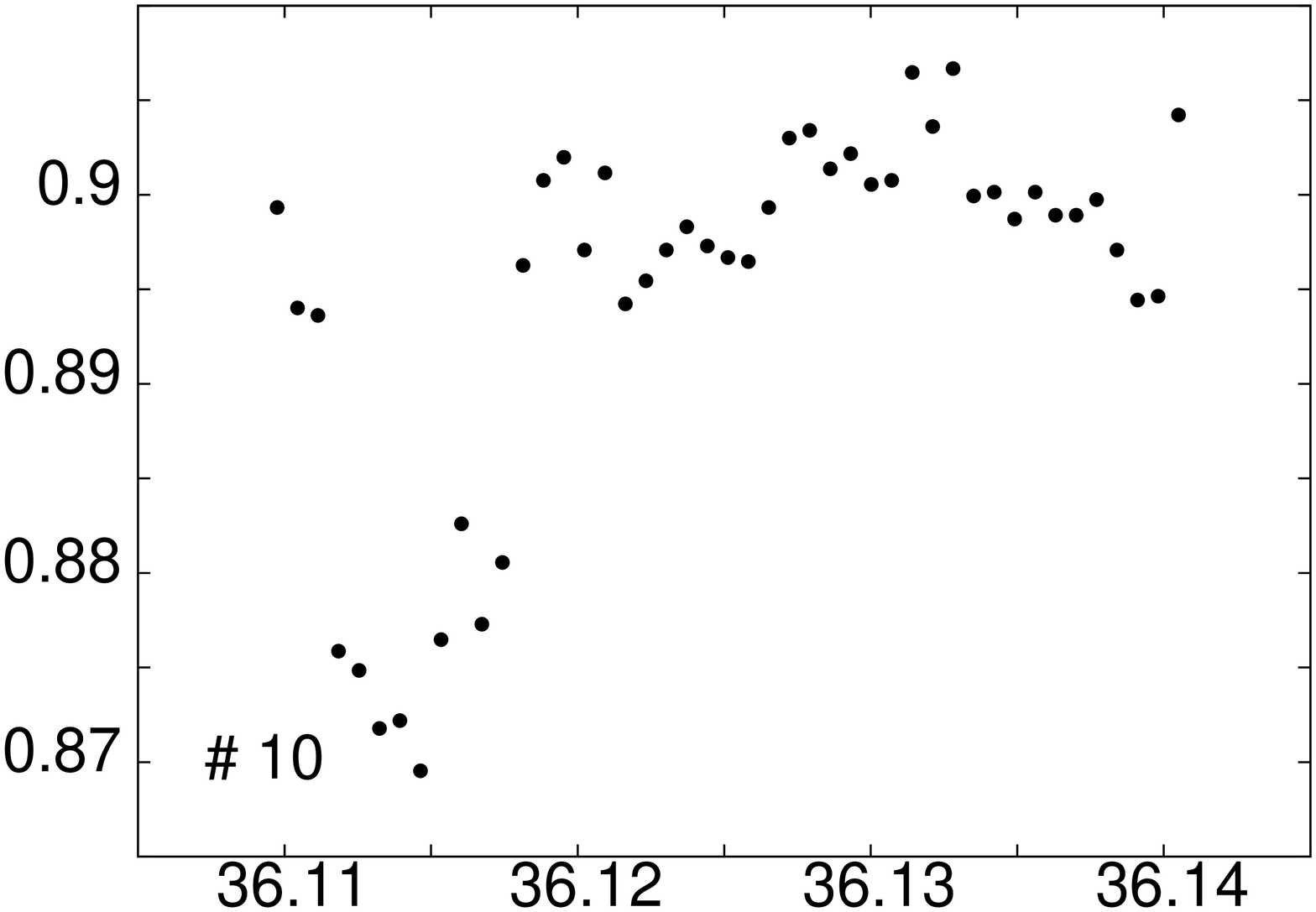}
\includegraphics[height=42mm,angle=-90]{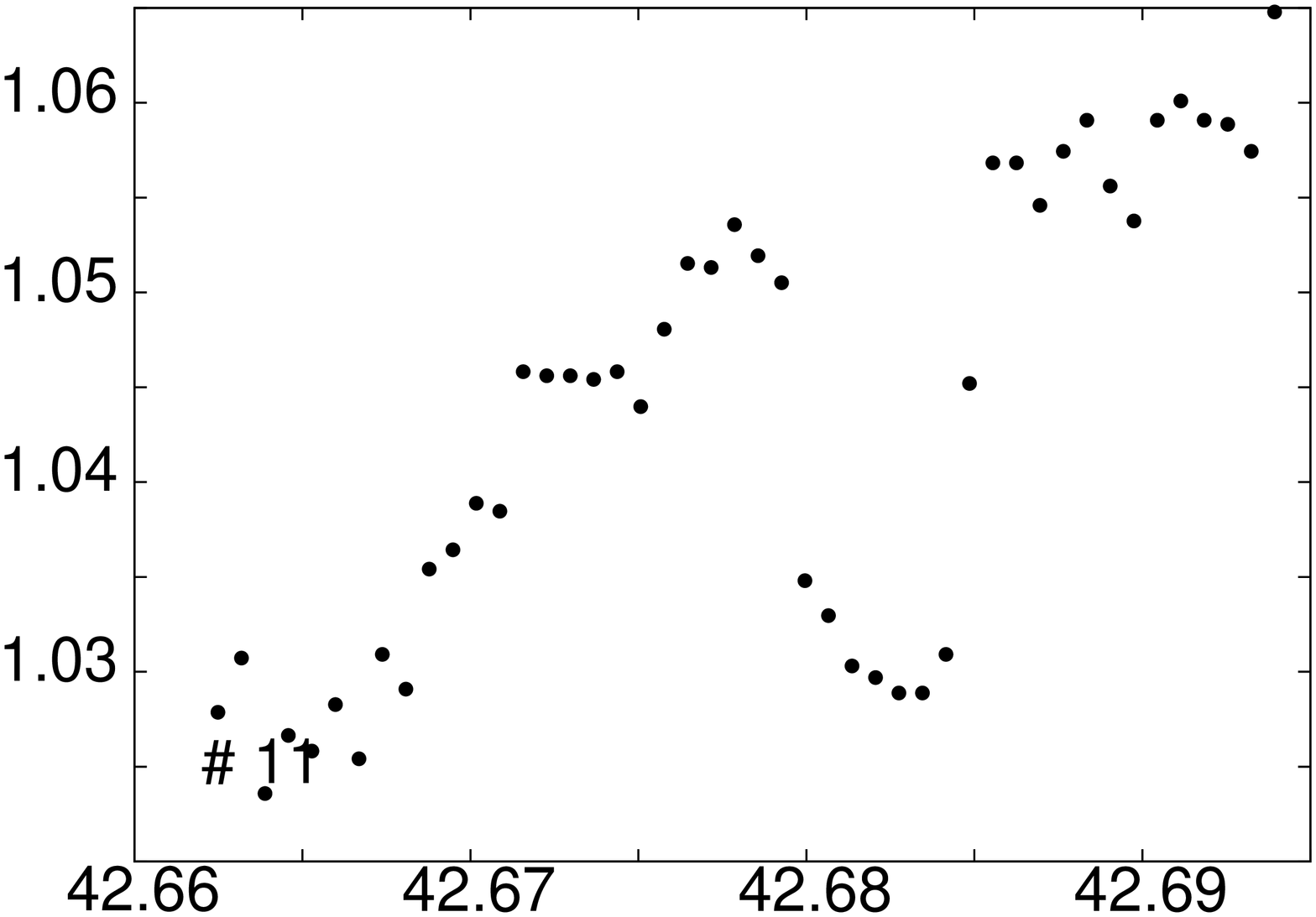}
\includegraphics[height=42mm,angle=-90]{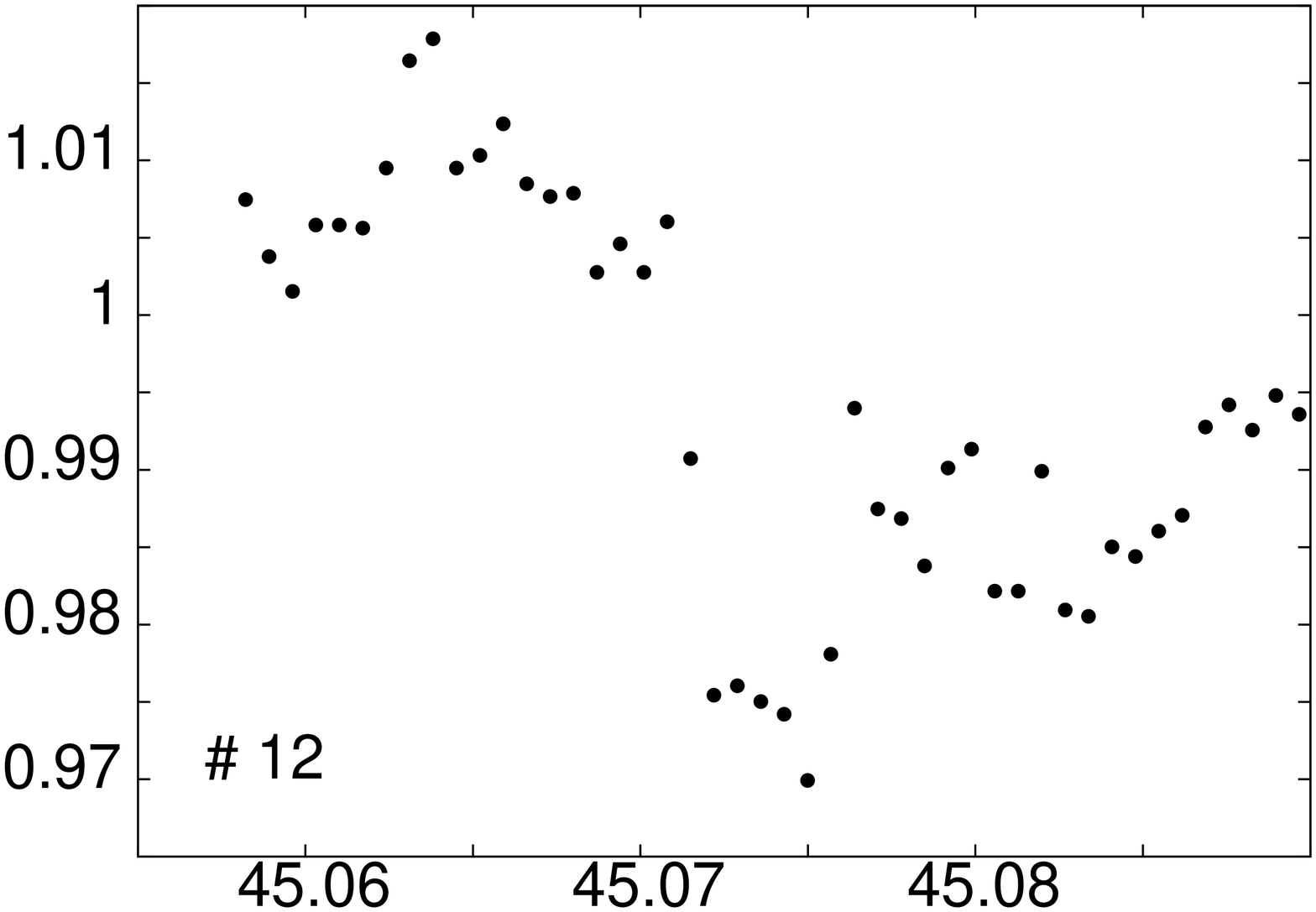}\\ 
\includegraphics[height=42mm,angle=-90]{figures/t12.eps}
\includegraphics[height=42mm,angle=-90]{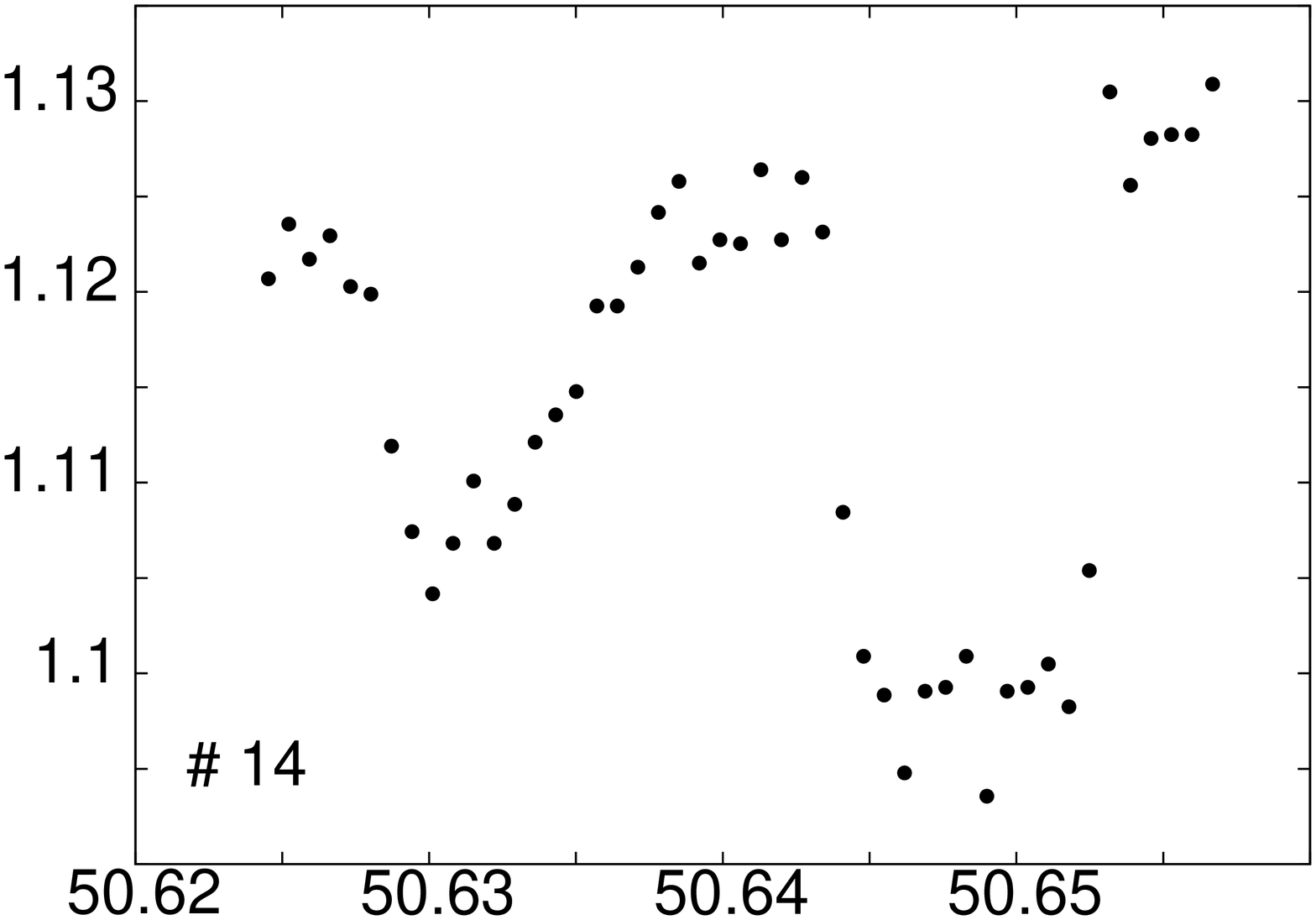}
\includegraphics[height=42mm,angle=-90]{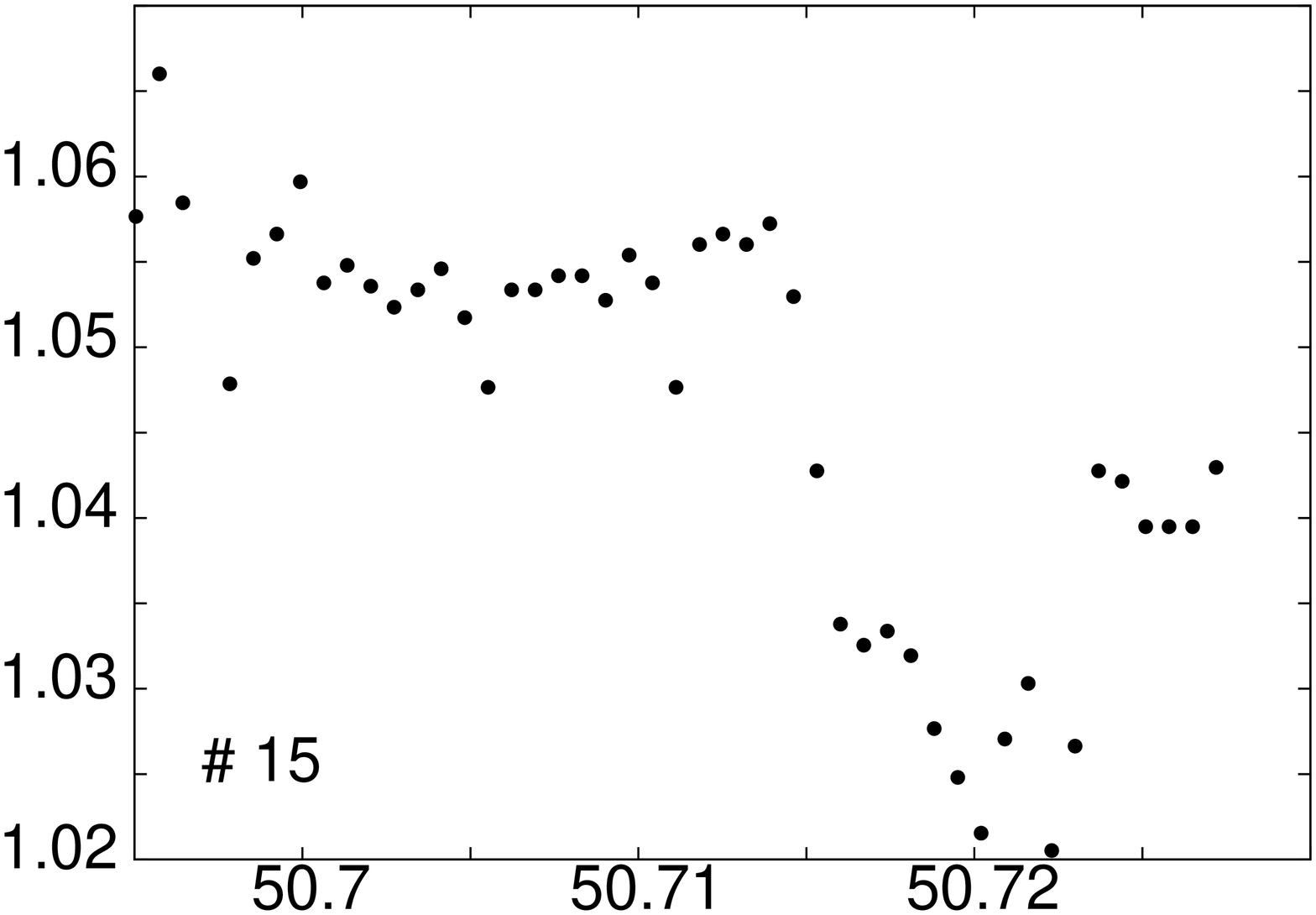}
\includegraphics[height=42mm,angle=-90]{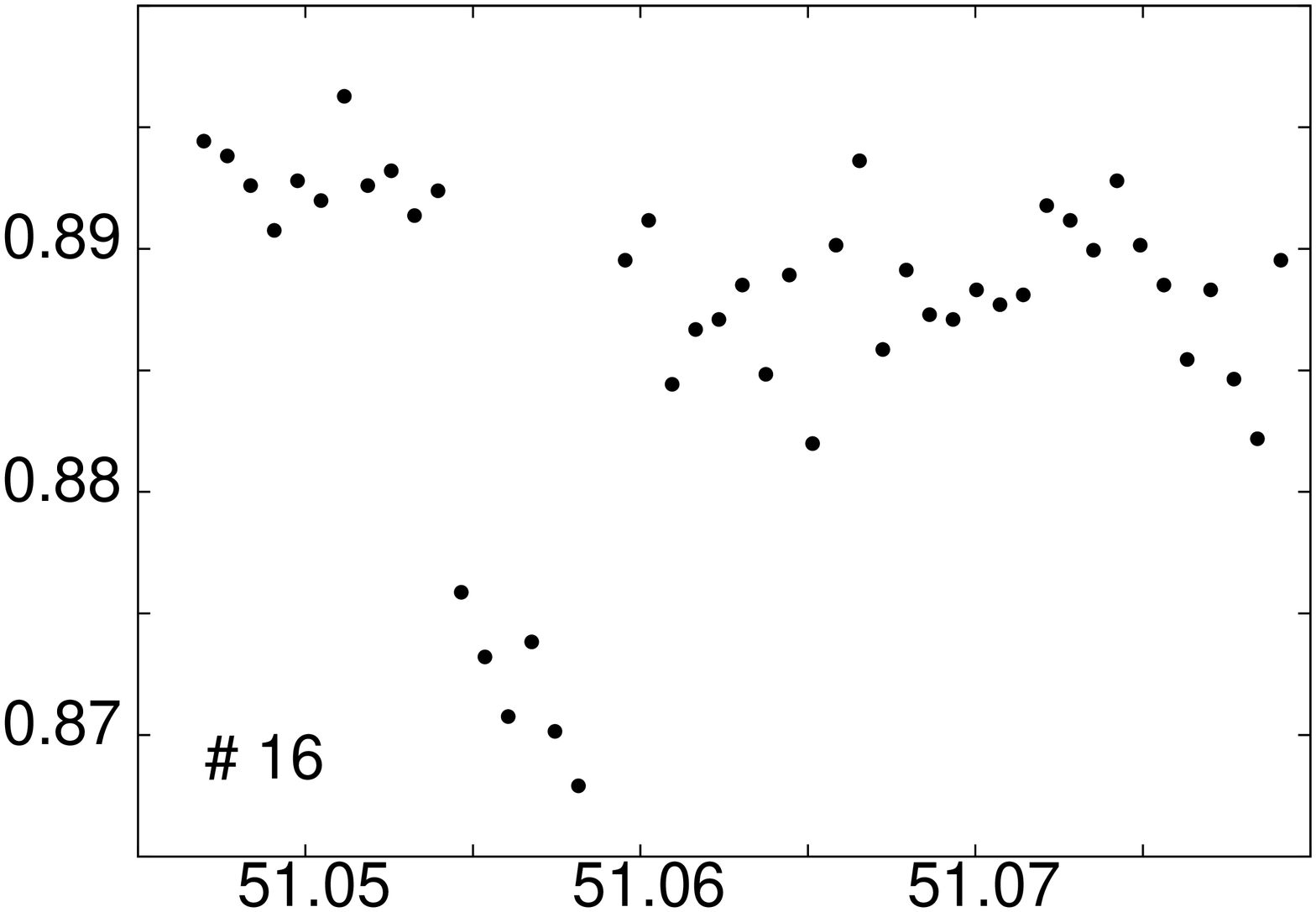}\\
\includegraphics[height=42mm,angle=-90]{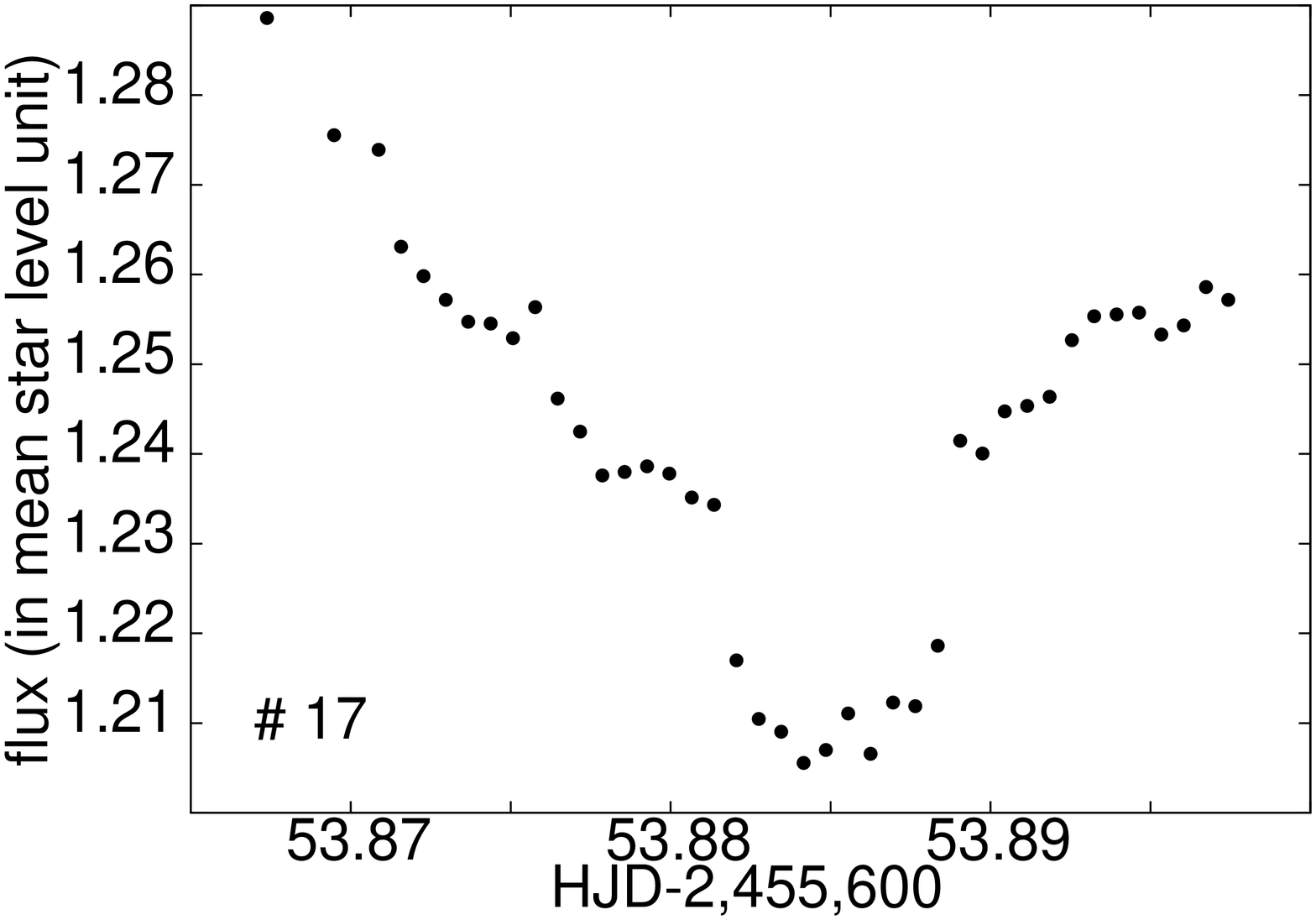}
\includegraphics[height=42mm,angle=-90]{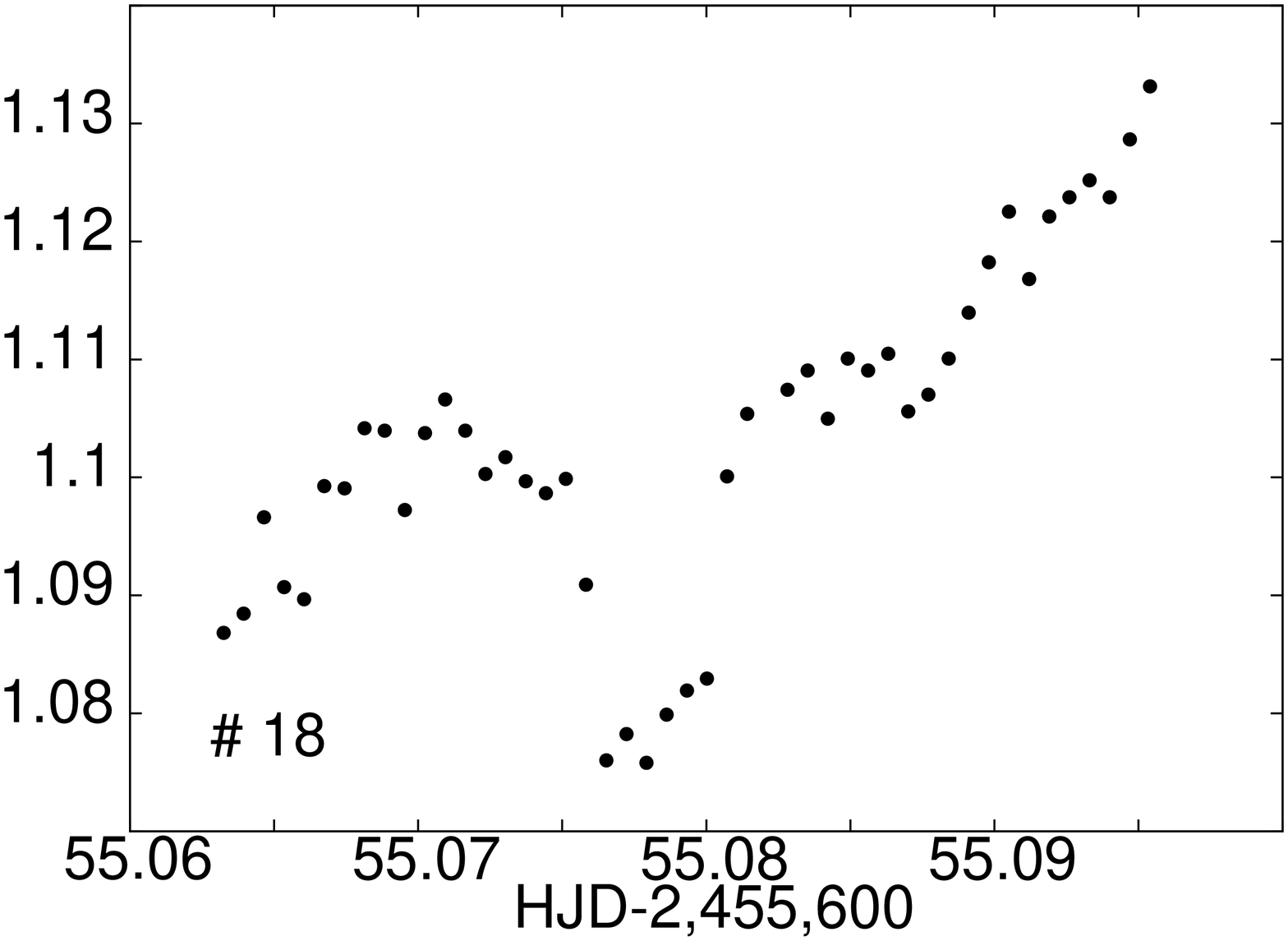} 
\includegraphics[height=42mm,angle=-90]{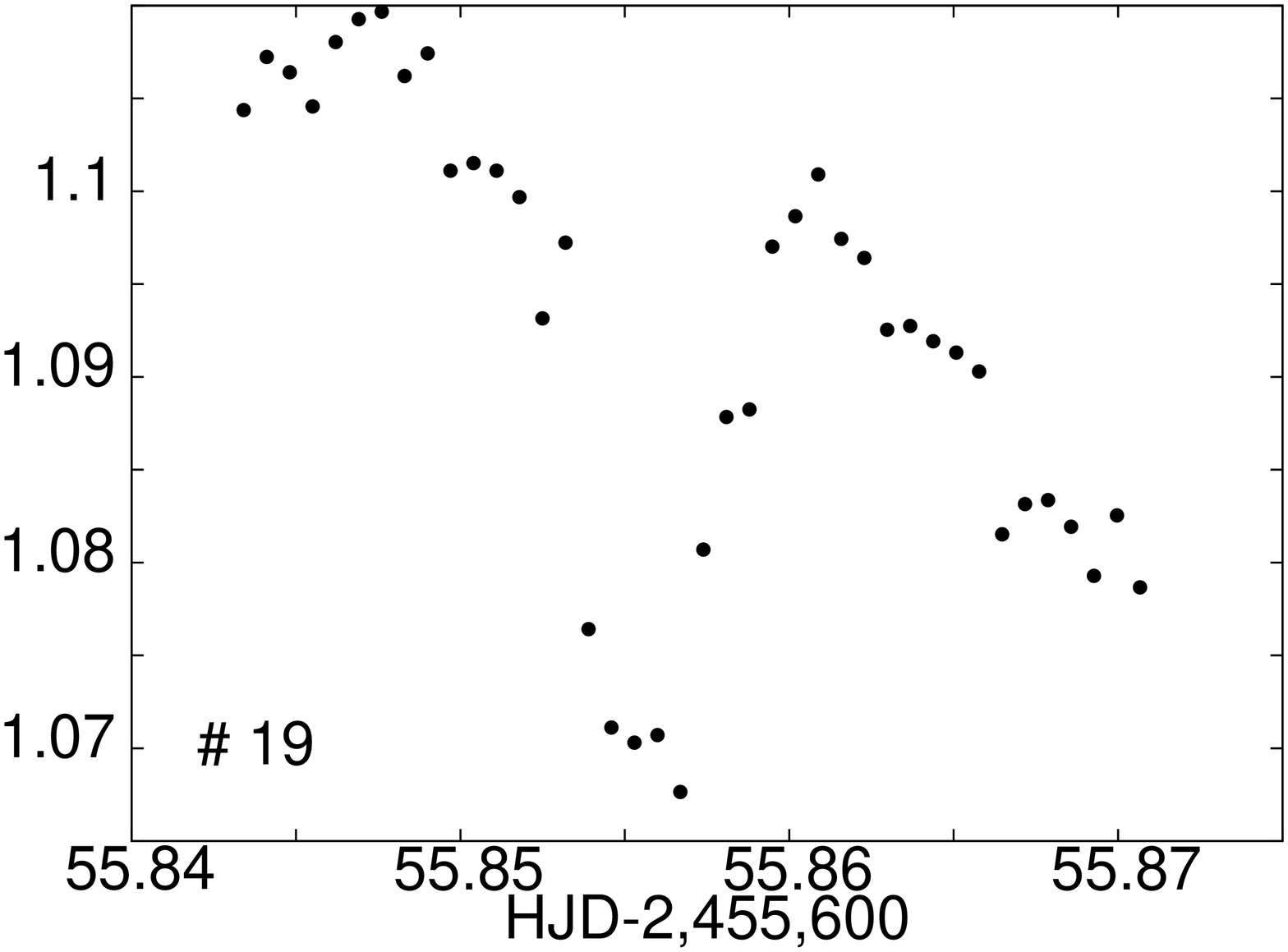}
\includegraphics[height=42mm,angle=-90]{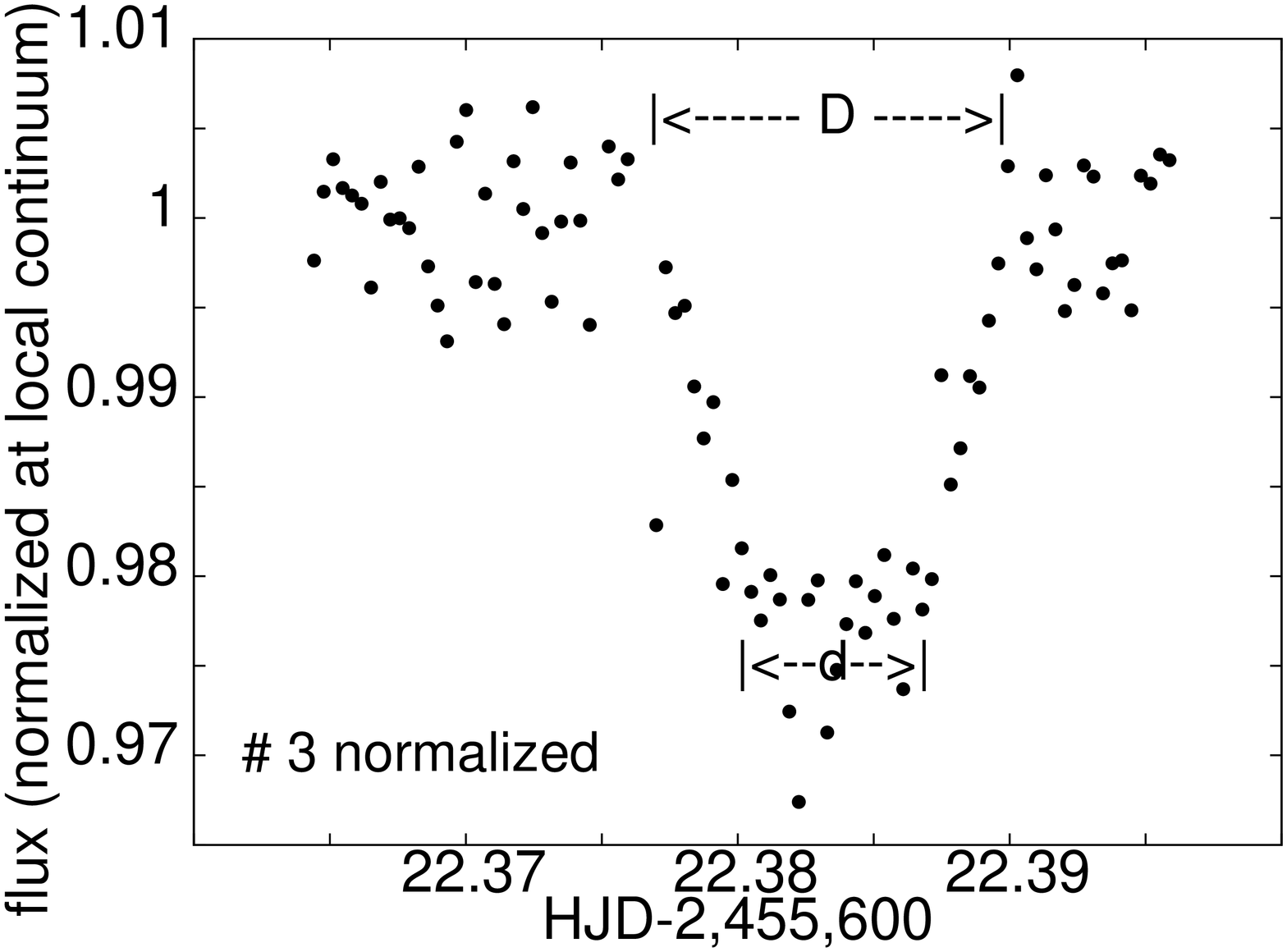}
\end{tabular}}
\caption{The fragment of the 2011 {\it MOST\/} light curve (upper large panel) with occultations 
$\#2,3\&4$, shown in detail on bottom panels along with all occultations detected 
by {\it MOST\/}. The numbers of occultations are given in the bottom left corner of each figure. 
The flux is left in normalized flux unit of the star mean brightness, i.e. the same as in 
Figure~\ref{Fig.1}. Occultations no. 1-8 were observed with 30~s integrations, occultations 
no. 9-19 with the 60~s integrations. The latest small panel shows the occultation \#3 normalized 
to the local brightness variations, with the horizontal 
arrows giving the definitions of the outer $D$ and inner $d$ contact duration times.
}
\label{Fig.4}
\end{figure*}

\begin{table}
\caption{Basic properties of occultations (Fig.~\ref{Fig.4}): 
the central dip moments $hjd_{min}$=$HJD-2,455,600$ 
estimated from mid-times of inner contacts, 
the dip depths related to the continuum flux assumed as unity
(see Sec.~\ref{eclipses}), the outer $D$ and inner $d$ contact durations in days 
(for description see also the last panel in Fig.~\ref{Fig.4}). 
The errors of first two quantities are given in parenthesis. 
Typical uncertainties of $D$ and $d$ are about 0.0003~d (0.5~min)}.
\begin{tabular}{c c c c}
\hline
no.     & $hjd_{min}$ [d]         & depth~[\%] & $D$~[d]~~~~~$d$~[d] \\ \hline
1              & 18.07835(17)     &  2.75(59)  & 0.0061~~~~~0.0040   \\
2              & 20.98028(23)     &  2.78(48)  & 0.0158~~~~~0.0109   \\ 
3              & 22.38364(51)     &  2.13(38)  & 0.0137~~~~~0.0061   \\     
4              & 23.80000(44)     &  3.03(41)  & 0.0134~~~~~0.0102   \\
5              & 25.55569(36)     &  2.57(46)  & 0.0093~~~~~0.0066   \\
6              & 26.04961(9)      &  2.67(43)  & 0.0072~~~~~0.0045   \\
7              & 26.97578(19)     &  2.63(56)  & 0.0055~~~~~0.0033   \\ 
8              & 30.05245(15)     &  2.42(23)  & 0.0039~~~~~0.0018   \\
9              & 32.74260(30)     &  2.56(32)  & 0.0078~~~~~0.0043   \\ 
10             & 36.11464(38)     &  2.56(31)  & 0.0082~~~~~0.0042   \\
11             & 42.68241(21)     &  2.45(11)  & 0.0068~~~~~0.0033   \\ 
12             & 45.07359(85)     &  2.29(14)  & 0.0063~~~~~0.0030   \\
13             & 48.02946(35)     &  1.87(35)  & 0.0119~~~~~0.0076   \\
14             & 50.64864(33)     &  2.49(20)  & 0.0102~~~~~0.0065   \\   
15             & 50.71951(21)     &  2.17(34)  & 0.0099~~~~~0.0063   \\
16             & 51.05640(41)     &  2.14(24)  & 0.0064~~~~~0.0029   \\
17             & 53.88521(7)      &  2.31(18)  & 0.0077~~~~~0.0049   \\  
18             & 55.07829(33)     &  2.20(23)  & 0.0067~~~~~0.0031   \\
19             & 55.85564(28)     &  2.80(10)  & 0.0072~~~~~0.0022   \\ \hline  
\end{tabular}
\label{Tab.1}
\end{table}

\subsection{The discovery}
\label{discovery}

Because of TW~Hya's location slightly
outside of the satellite Continuous Visibility Zone, its {\it MOST\/} observations
must be interrupted during each orbit. During the 2011 run, the star
was observed for typically 40 -- 50 minutes during each 101 minute orbit, which
makes the current observations very well suited for detection of short-lasting events. 
This good time coverage (see in Figure~\ref{Fig.1}, and also in 
Section~\ref{observations})
is important to what we detected and should be contrasted with 
the previous observations which lasted typically 
20 -- 40 minutes per orbit. 
Additionally, to permit multiplexing with other targets, the previous 
runs included further time limitations with observations of TW~Hya during only 
every second or third satellite orbit.
The crucial advantage of the longer than ever before satellite pointing  
during the 2011 run has led to a discovery of an entirely new phenomenon: 
the light curve showed 19 short, well defined dips (Figure~\ref{Fig.4}) which 
we simply call ``occultations''. 
As an occultation, we define a flux decrease 
to a flat or nearly flat bottom and then an increase 
by the same flux amount which appears similar to eclipses 
in detached, non-interacting eclipsing binary stars. 
We require them to be clearly distinguishable from the 
variety of other quasi-periodic and stochastic, but smooth light variations visible 
in single-orbit data, even if they take place close in time to 
the occultations; this point is illustrated in Figure~\ref{Fig.4}.

\begin{figure*}
\centerline{%
\begin{tabular}{c@{\hspace{0pt}}c}
\includegraphics[height=85mm,angle=-90]{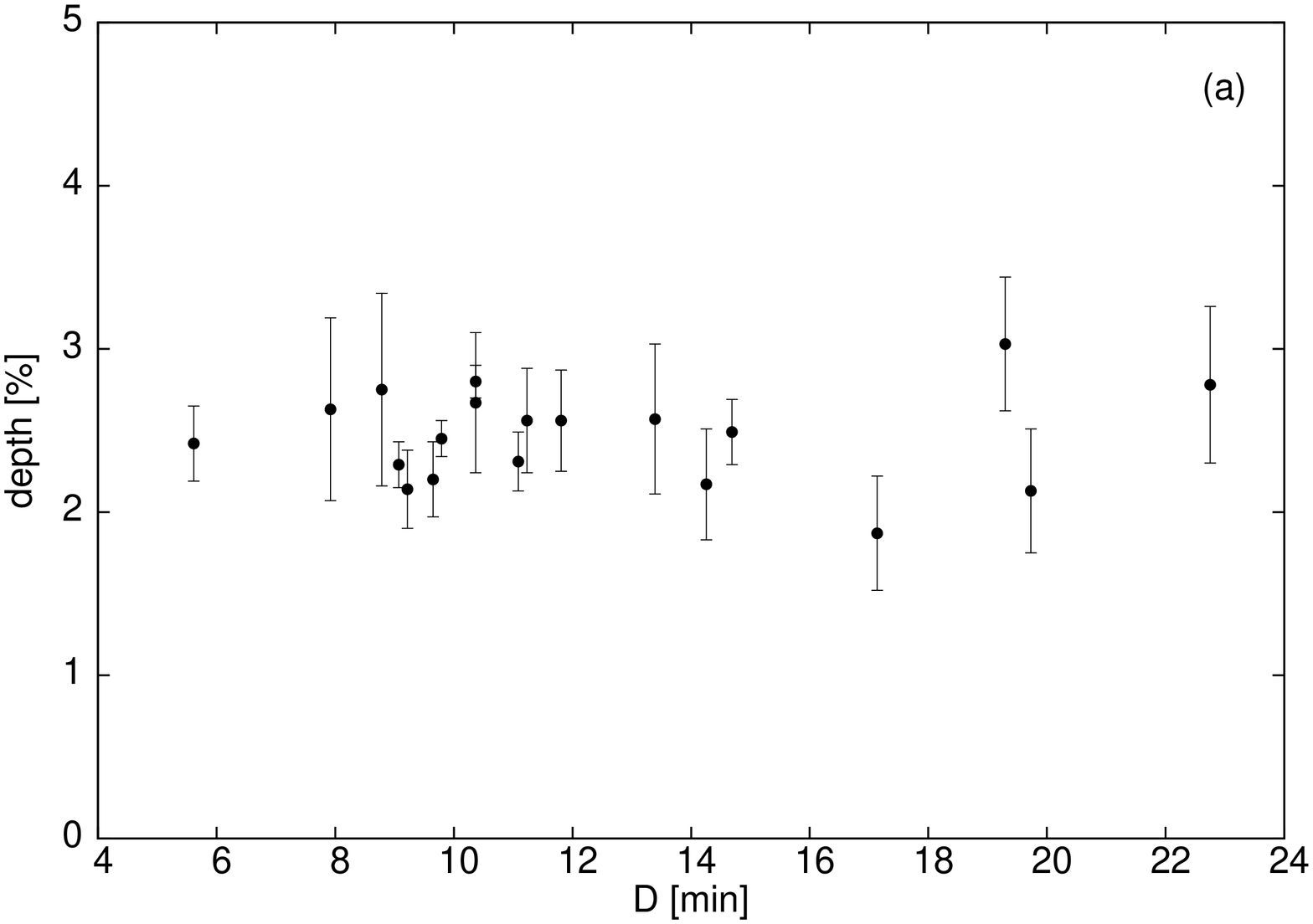}&
\includegraphics[height=85mm,angle=-90]{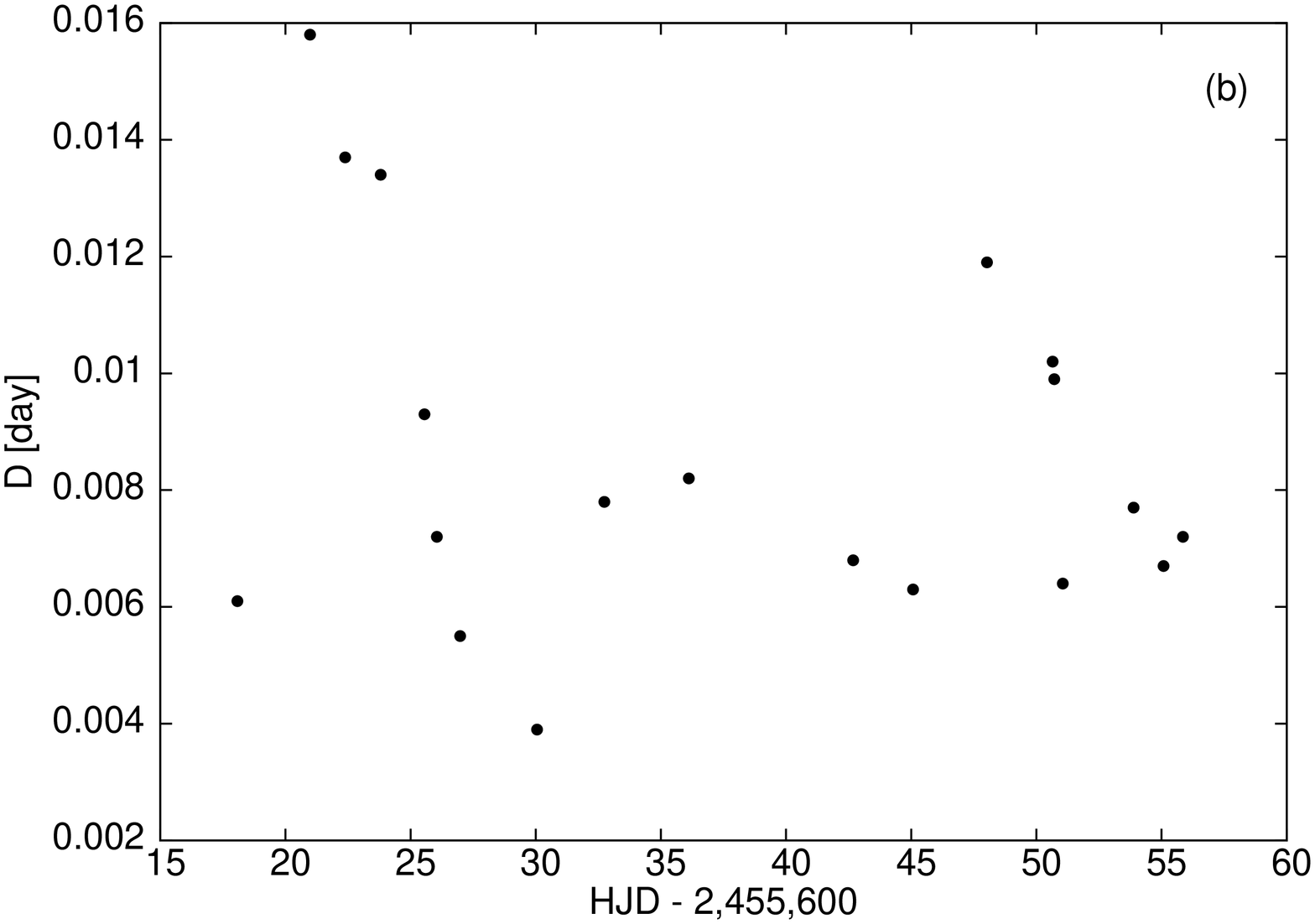}\\
\includegraphics[height=85mm,angle=-90]{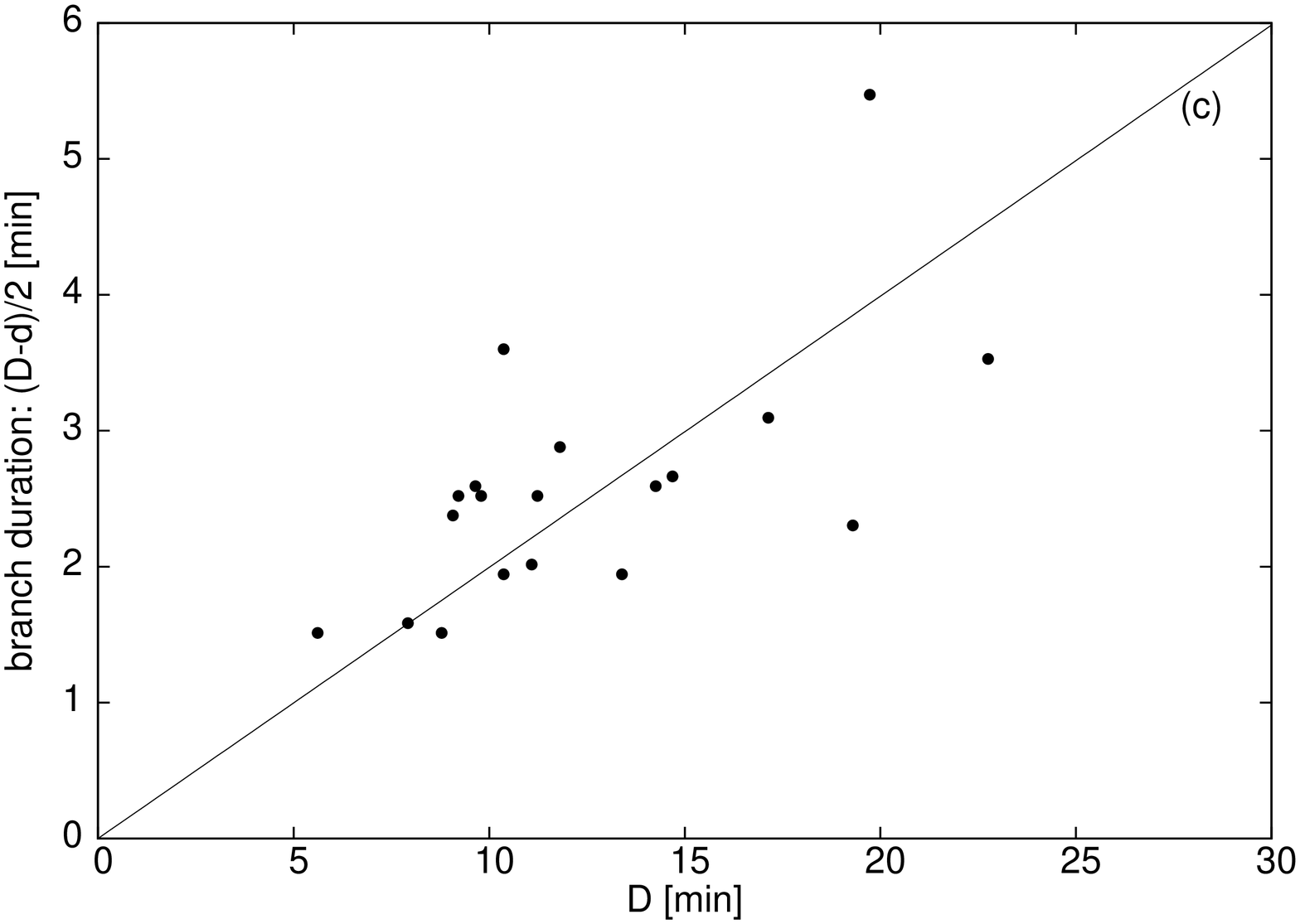}&
\includegraphics[height=85mm,angle=-90]{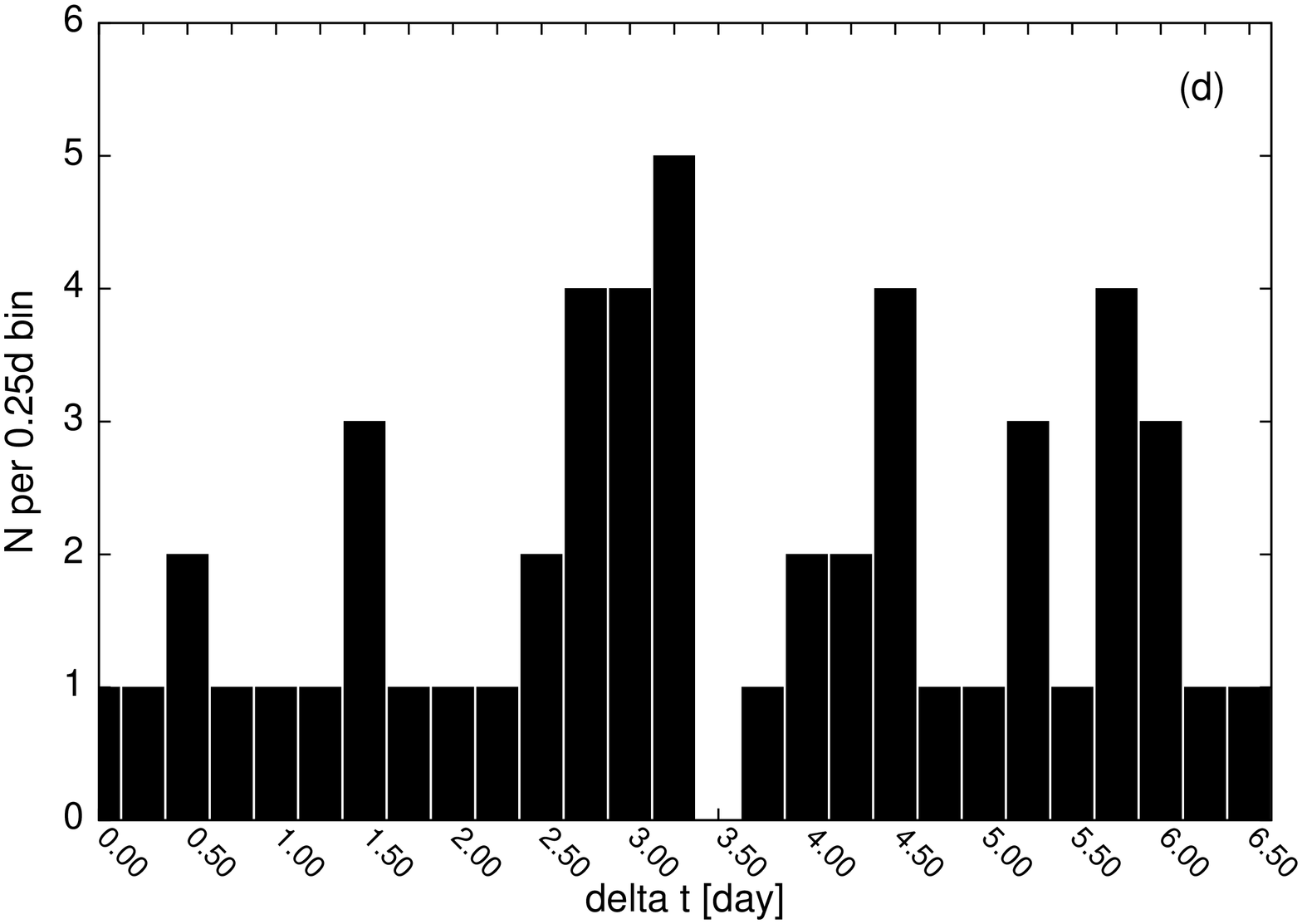}
\end{tabular}}
\caption{The figure presents relations between 
various characteristics of
the occultations: 
(a) The relation between the occultation durations $D$ 
and their relative depths (in percent),
(b) The distribution of outer contact durations $D$ in time, 
(c) The relation (a straight-line fit) between the durations $D$ and the branch 
durations $(D-d)/2$: $(D-d)/2=0.199(14)D$, and 
(d) The histogram of spacings between all available pairs of occultations.
}
\label{Fig.5}
\end{figure*}

Table~\ref{Tab.1} gives the central mean moments of the minima $hjd_{min}$, 
their depths and the durations of intervals 
between the outer $D$ and the inner $d$ contacts. 
Figure~\ref{Fig.5} gives the essential characteristics of the occultations
such as relative depths, durations and spacings in time. 

The {\it MOST\/} observations are obtained in one photometric 
filter so no temperature information is available for 
interpretation of the phenomenon. 
Because we do not know what causes the occultations, we can give only
a purely heuristic description of their properties:
\begin{itemize}
\item
From the number of observed 19 occultations with 
the effective uninterrupted coverage of 16.92~days 
(see Section~\ref{observations}), we derive
the mean rate of occurrence of 1.12 occultation per day.
\item
Two additional shallow occultations may have been seen as well, 
but due to the breaks in temporal coverage it is hard to distinguish them from 
irregular variability intrinsic to the accretion effects. 
Two more occultations occurred too close to the ends of the {\it MOST\/} 
orbits. These four, ambiguous events are not considered here; 
their inclusion would increase the frequency to 1.36 occultation per day.
\item
Most of the events are flat-bottomed and look as though they were caused 
by total occultations. 
\item
To characterize the occultations in a uniform way, we removed the smooth brightness 
variations using low-order polynomials fitted to the neighbouring continuum. 
Their depths measured relatively to the local 
flux continuum normalized to unity are 
surprisingly similar and range between 1.87\% and 3.03\% 
(see Table~\ref{Tab.1} and in Figure~\ref{Fig.5}, panel "a"). 
\item 
In Figure~\ref{Fig.5}, panel "b", 
we show the distribution of occultation durations. The median value
of the full durations is  $D = 0.0077\pm0.0031$~d ($11.1\pm4.5$~min) while 
the median duration of the total occultation is  $d = 0.0043\pm0.0025$~d
($6.2\pm3.6$~min).  The branches, i.e $(D - d)/2$ are very short, on the
average lasting $0.00175\pm0.0006$~d ($2.5\pm0.9$~min).
\item  
We found four occultations (\#2, 3, 4, 13) lasting twice 
longer than the median duration time (see Table~\ref{Tab.1}).  
Their branches also last longer in the same proportion. 
The relation $(D-d)/2=0.199(14) \times D$ between the branch durations $(D-d)/2$ 
and the full durations $D$ is shown in Figure~\ref{Fig.5}, panel "c".
\item
We do not see any single occultation lasting shorter than 5~minutes,
but this may be partly due to the spacing of observations of one minute.
Also, due to the specific format of {\it MOST\/} data (see in Figure~\ref{Fig.4}) 
dips lasting longer than about $3/4$ of a single {\it MOST\/} orbit length (i.e. $35$~min) 
could 
be undetected; such longer time scales (1-2~h) 
can be investigated with ground based telescopes. 
\item
Regularity in the distribution of the occultations in time
is difficult to characterize because of the breaks in the temporal coverage. 
Yet, we note that spacings between two pairs of occultations,  
\#8 and \#9 (2.69015~d) as well as \#13 and \#15 (2.69005~d) 
are identical within the measurement errors ($\approx 0.0003$~d, Tab~\ref{Tab.1}). 
However, these four occultations cannot be phased with one linear relation.
In contrast to the pair \#8 and \#9, 
the occultations forming the pair \#13 and \#15 look very similar, in that
their durations 
$D$ are comparable and they show the same characteristic shape of light 
variation. 
We also note the very similar shapes, depths and durations of the occultations 
\#2 and \#4, separated by a similar amount of time (2.81972~d).
\item
The issue of the regularity of occurrence appears to be crucial in
any attempts to find the process producing the occultations. But, even
for the time series as uniform as our observations, they have unavoidable
gaps in the coverage caused by the location of TW~Hya outside of the satellite 
Continuous Viewing Zone.
In order to minimize the effects of breaks in temporal coverage, 
we calculated spacings between all possible pairs of occultations 
and the resulting values were binned into 0.25~d intervals 
(see Figure~\ref{Fig.5}, panel "d"). 
There exists a very well defined primary broad peak, 
around 2.5 to 3.3 days while the 5.75~d peak may be related to this feature
for double spacing.  Some secondary 
concentrations may be present at $\sim1.5$~d and $\sim4.5$~d but their
significance is low. 
We note that all these features correspond to time scales much shorter
than the duration of the {\it MOST\/} run and are not 
affected by its finite duration.

We tested the histogram of spacings between occultations 
using the Kolmogorov-Smirnov test which compares the
respective cumulative distributions. The formal significance 
for the distribution as in Figure~\ref{Fig.5} (panel "d") 
being identical to a uniform distribution is only 0.001.

\end{itemize}

\subsection{Have the occultations been seen before?}
\label{history}

The large number of occultation events in the 2011 observations of
TW~Hya leads to an obvious question: Have they been seen before and
we simply overlooked them in our 2007, 2008 and 2009 {\it MOST\/} observations? 
The accuracy of 2007 data was too small to allow for their detection. 
Although the 2008 and 2009 runs lasted 46.7 and 40.3 days, their effective 
strictly continuous time coverages was only 6.96 and 4.62~d, respectively. 
In this respect, the 2011 run with the effective coverage of 16.92~days
was by far more conducive for detection of the occultations. 
Nevertheless, taking into account the effective coverage time ratios, 
we should have seen a few occultations (i.e. 8 and 5, respectively) 
in the older data.

\begin{figure*}
\centerline{%
\begin{tabular}{c@{\hspace{20pt}}c@{\hspace{20pt}}c}
\includegraphics[height=55mm,angle=-90]{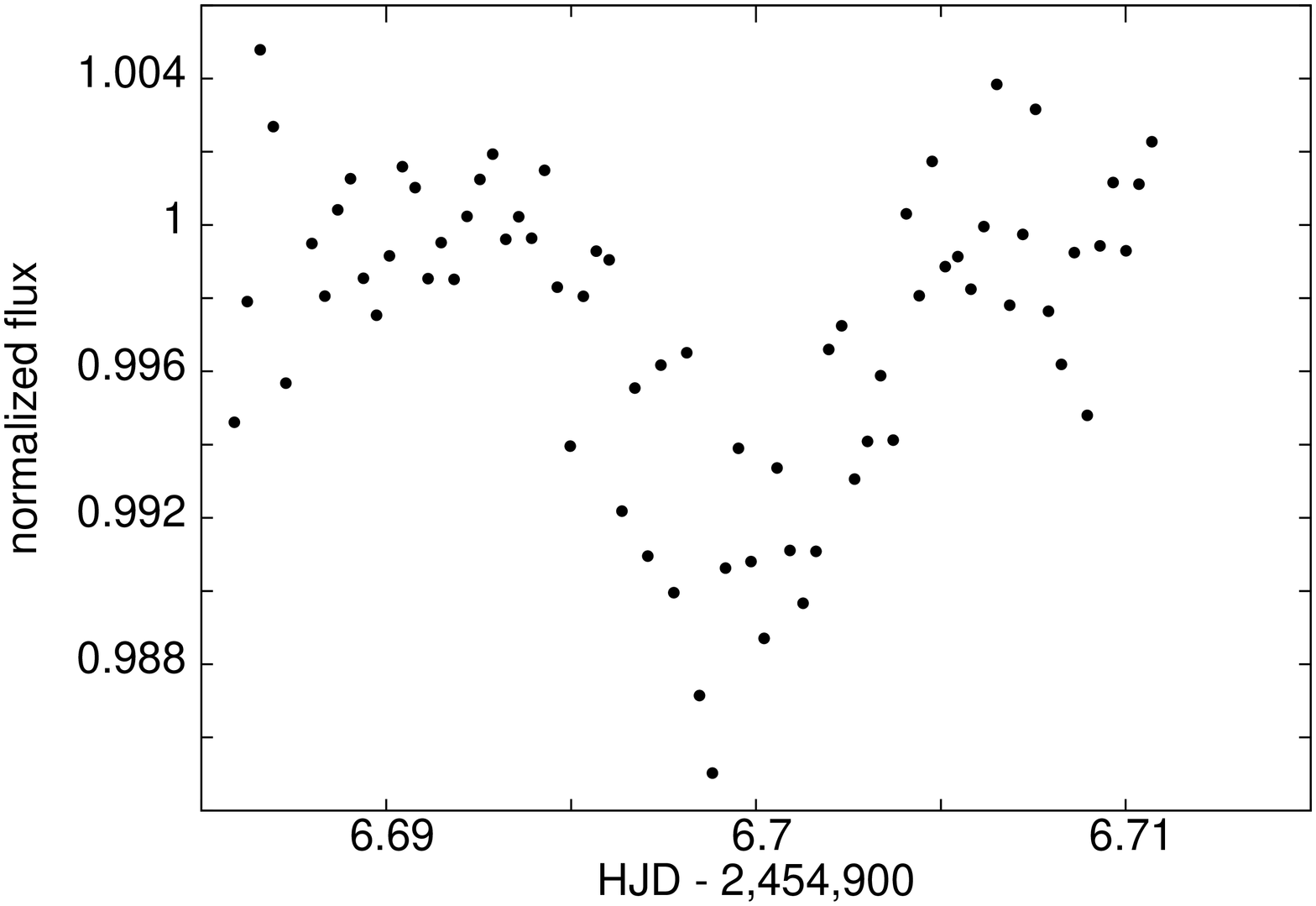}
\includegraphics[height=55mm,angle=-90]{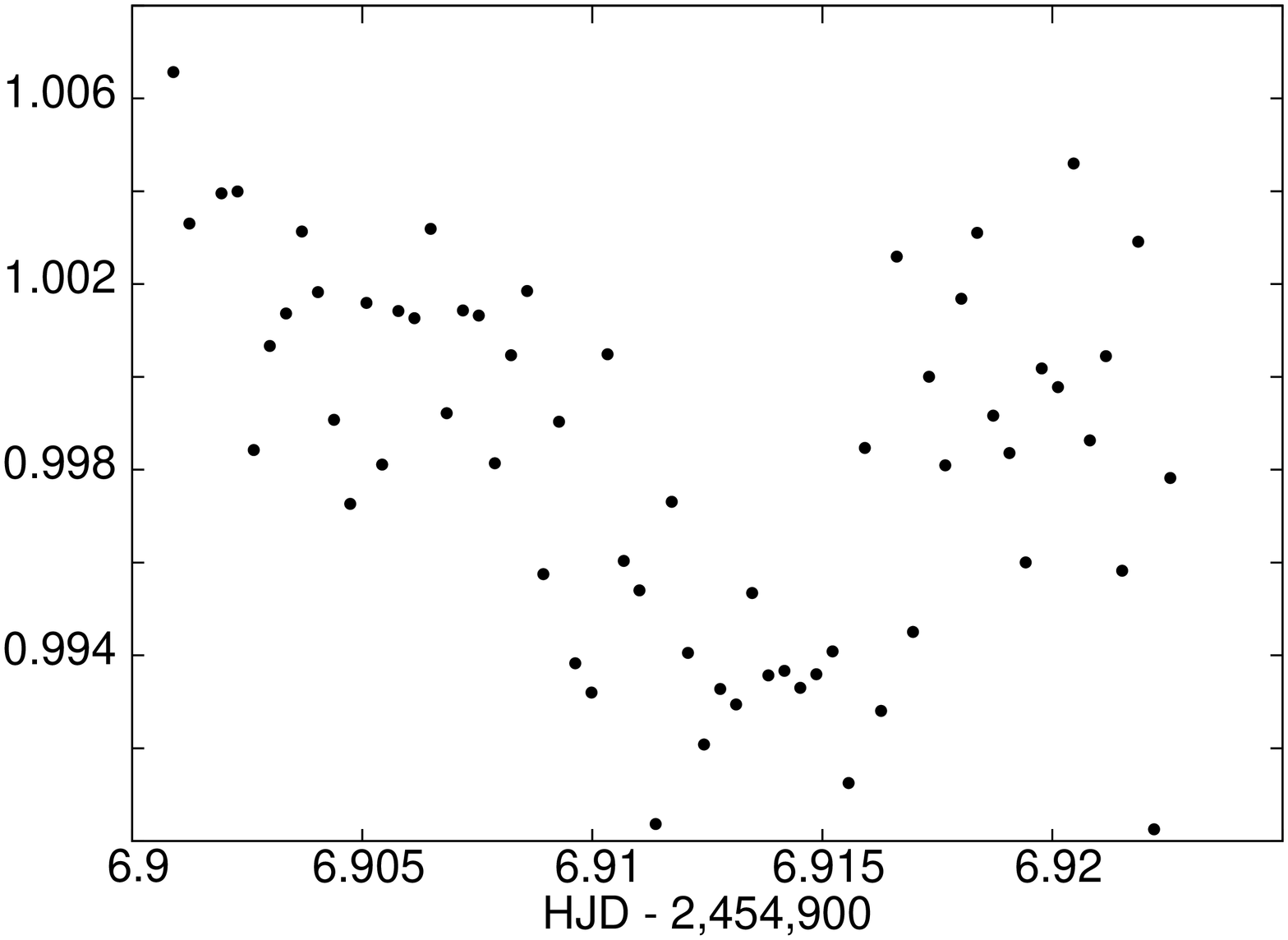}
\includegraphics[height=55mm,angle=-90]{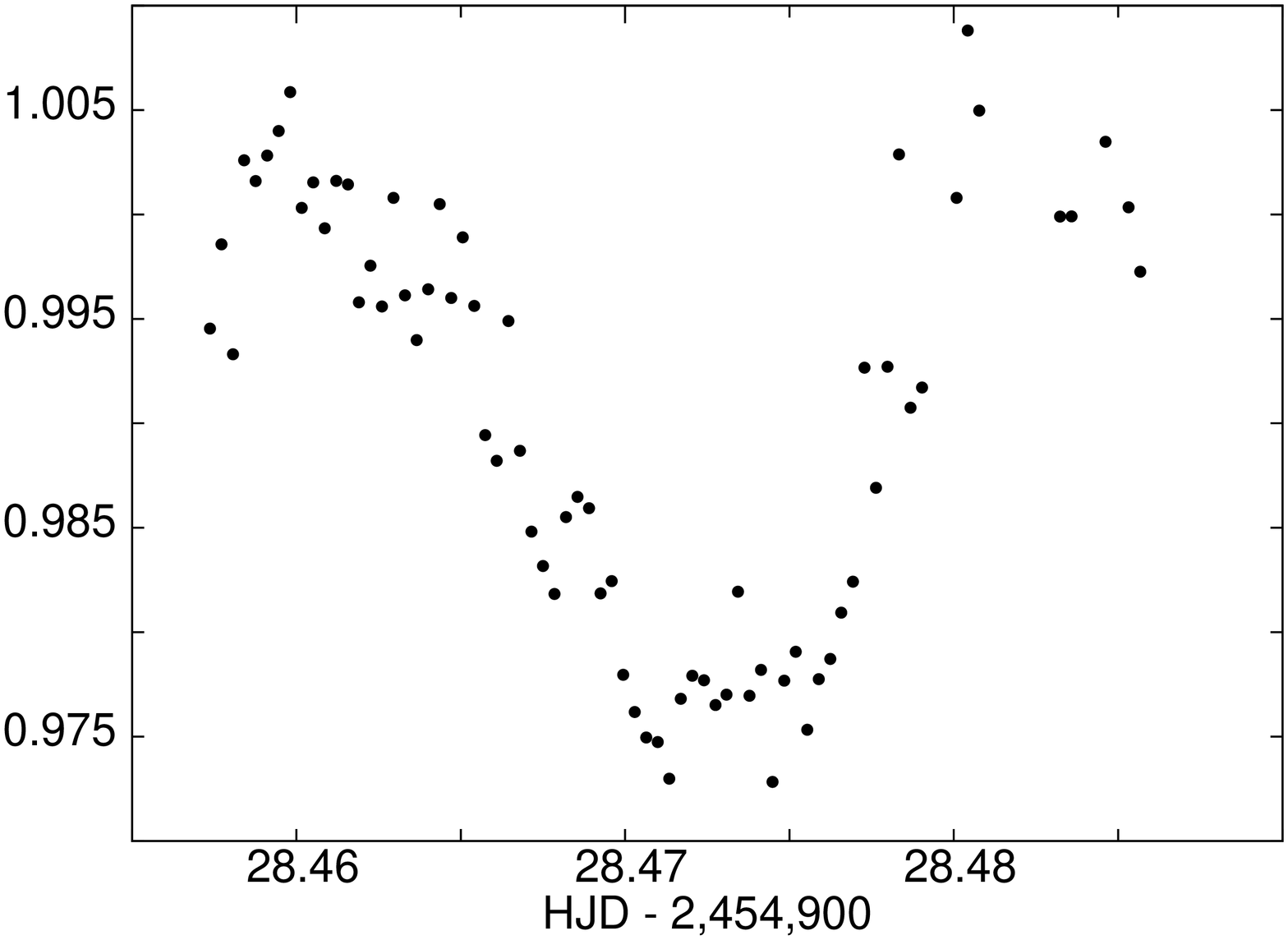}
\end{tabular}}
\caption{Possible occultations in the {\it MOST\/} 2009 light curve 
of TW~Hya, i.e. those obtained two years before the discovery observations reported here. 
Only the last two events fully meet definition of an occultation as given in Section~\ref{discovery}.}
\label{Fig.6}
\end{figure*}

Although several brightness drops in the 2008 data could potentially
be similar to the occultations discussed here, none has such a well
defined shape as those observed in 2011.
Three possible occultations appear to exist in the 2009 {\it MOST\/} 
observations at $HJD-2,454,900 = 6.700(1), 6.914(1)$ and $28.473(1)$ 
(see Figure~\ref{Fig.6}). 
Their depths are 1.0\%, 0.7\% and 2.3\%, respectively, but only the last 
event is as well defined as any of the 2011 occultations. 
We infer that perhaps due to different conditions producing the occultations, 
their shape, depth and occurrence rate can evolve from season to season.

Other T~Tauri stars may have also shown the occultations. 
The rapid brightness drops, but of very different depths 
and durations were noted in 
high speed photometry of the Classical T~Tauri-type star DD~Ser, 
\citet{verlyuk} where much deeper, 0.3-0.7~mag 
dips lasted only 4-5~sec. A single 1.2~mag, very brief (1.7 sec) 
dip was also observed in AB~Aur, a Herbig~Ae star. For the latter
star the inclination angle of the accretion disc is 
$29.8\pm1.3$~deg (\citealt{hashimoto}), which is similar to the 
nearly pole-on disk in TW~Hya.

\subsection{What is occulted and what is causing the occultations?} 
\label{speculations}

\citet{debes} suggest that a planetary formation process is occurring 
in the TW~Hya accretion disc. However, because of the pole-on orientation
of the rotation axis of the star and of the inner disc visible in infrared light 
\citep{krist00, potter05}, one would not expect to observe planets 
forming close to the disc plane and passing in front of the star.
On the other hand, the range of planetary-orbit inclinations in young stars may
be surprisingly large. We note the case of the weak-lined 
T~Tauri-type star CVSO~30 which was shown by \citet{vanEyken} to have 
a possible planet transiting for $i=62\pm4$~deg.

Although the occultation depths of 2-3\% in Table~\ref{Tab.1} are similar to those caused 
by transits of giant planets, the short branch durations 
of about 2 -- 3 minutes and the semi-regular incidence of the occultations
indicate that the occulted source is not the star itself but must have dimensions of
a sizeable fraction of the stellar radius. 
If the occultations were due to the transits in front of TW~Hya, the occulting bodies
would have to be large, about 0.15 of the diameter of the star. 
Moreover, even if a hypothetical planet would orbit the star on polar orbit 
with a semi-major axis similar to the inner accretion disc radius of 12~R$_{\odot}$ 
\citep{eisner06}, the expected transit duration times 
for the 0.4-0.7$M_{\odot}$ star \citep{vacca2011} would be about 2-3 hours.

Through process of elimination, we conclude that the occulted sources are most likely 
hot regions of small, yet finite dimensions. 
These could be the hot spots localized at footprints of 
accretion tongues created through RT instabilities and/or sources of numerous, 
strong emission lines observed in a spectrum of the star. 
The typical angular dimensions of the hot regions, as seen by the occulting body 
must be very small, roughly $\simeq 0.0006$~rad (0.03~deg), for the assumed 
2.5 minutes of the branch duration and 3 days for their characteristic reappearance 
time indicated by the main broad peak in Figure~\ref{Fig.5} (panel "d");
Presumably the inner disc and the stellar rotational frequencies comprise the most 
natural ''clocks" determining temporal occurrence of the occultations.\newline
We see no dependence between the eclipse depths and their durations 
(Figure~\ref{Fig.5}, panel "a") so that the eclipsed sources probably 
had very similar sizes and brightnesses during the entire 2011 {\it MOST\/} run.
If these sources would have an additional freedom to move, 
as predicted for hot spots created through RT 
instabilities rotating around the star with the inner disc rotational frequency, 
no strict periodicity would be expected between the eclipse events. 
This scenario could explain why the characteristic reappearance time of the 
occultations of 2.6-3.2~d (Figure~\ref{Fig.5}, panel "d") 
is shorter than the stable 4.18~d main signal; it could also 
explain the occurrence of two pairs of occultations 
separated by the same amount of time (2.69~d) 
and the third one separated by 2.81~d, which consists of 
dips with very similar shapes and duration times 
(see in Fig.~\ref{Fig.4} and in Section~\ref{discovery}).

Similarly to speculations on the nature of the occulted sources,
we can offer only speculations on the nature of the occulting bodies. 
In Figure~\ref{Fig.5}, panel "c", we note that the branch 
and eclipse durations are positively 
correlated. This is a very important feature which may be interpreted 
as indicating several occulting bodies 
positioned at different distances from the star. 
These could be free-floating dark clumps, perhaps elements of TW~Hya's Oort cloud, 
representing a more advanced stage of ''dusty traps" suggested on the basis of ALMA 
observations of Oph~IRS-48 by \citet{vanderMarel}. 
But they could be also small optically thick plasma 
condensations (if not the accretion stream itself) 
levitating in a magnetic field of the steady accretion funnel 
encircling the stellar-disc magnetosphere and acting as natural 
screens for the hot spots created at low or moderate latitudes of the star.
This scenario could explain the high occultation rate in 
2011, when the {\it moderately stable regime} of accretion operated 
in the star.

\section{Summary}
\label{summary}

The results of the 2011 {\it MOST\/} satellite observations of TW~Hya 
are exceptional when compared with the results from the 2007, 2008, and 2009
seasons. While the general light variations retained the  
general characteristics of flicker noise with amplitudes scaling as: 
$a \propto 1/\sqrt{f}$ (see also \citealt{ruc08,siwak11}), 
the 2011 season variations did not show any obvious 
period shortening of the oscillation features.
This time, the Fourier and wavelet spectra are dominated 
by a single almost stable oscillation
with a period of $4.18\pm0.25$ days. 
We propose that the dominant oscillation is due to rotational 
modulation produced by a single, large hot spot 
formed close to the magnetic pole, at the footprint of the accretion 
funnel which can originate only during the {\it stable} or {\it moderately stable regime} 
of accretion \citep{romanowa04,kulkarni09}. 
Within this framework, the fairly stable 4.18~d signal could represent the true 
rotational period of the star (see also \citealt{batalha02,kurosawa13}). 
It is not clear at this time how the 4.18~d periodicity relates 
to the previously observed 3.57~d spectroscopic period.

A new phenomenon has been detected in the light curve of TW~Hya consisting
of numerous relatively short-duration (10 -- 20~min), 2 -- 3\% deep drops in brightness which 
we call occultations. 
Their short branches, lasting typically about 2.5 minutes indicate finite dimensions 
of the occulted source(s). 
Although without any temperature information we are unable to firmly interpret the observed phenomenon, 
we suggest that these can be due to occultations of small hot spots on the star, created at moderate latitudes 
at the footprints of a few accretion tongues produced through the Rayleigh-Taylor instabilities. 
The obscuring body could be some dark, free-floating condensed clumps orbiting 
the star on highly inclined orbits beyond the magnetosphere or levitating in magnetic 
fields optically thick plasma condensations within the accretion funnel.
It is expected that multi-colour and spectral, high-cadence observations of TW Hya 
will bring crucial information about the temperature of the occulted regions to provide 
more firm interpretation for our discovery.

\section*{Acknowledgments}
MS is grateful for the Polish National Science Centre 
grant 2012/05/E/ST9/03915, fully supporting his research.\newline
The Natural Sciences and Engineering Research Council of
Canada supports the research of DBG, JMM, AFJM, and SMR.
Additional support for AFJM comes from FQRNT (Qu\'ebec).
RK is supported by the Canadian Space Agency and WWW is supported 
by the Austrian Science Funds (P22691-N16).\newline
This research has made use of NASA's Astrophysics Data System (ADS) 
Bibliographic Services.\newline
Special thanks are also due to an anonymous referee for highly useful suggestions
and comments on the previous version of the paper.

\end{document}